\documentclass[sigconf]{acmart}

\if 0 
\copyrightyear{2026}
\acmYear{2026}
\setcopyright{cc}
\setcctype{by}
\acmConference[WWW '26]{Proceedings of the ACM Web Conference 2026}{April 13--17, 2026}{Dubai, United Arab Emirates}
\acmBooktitle{Proceedings of the ACM Web Conference 2026 (WWW '26), April 13--17, 2026, Dubai, United Arab Emirates}
\acmPrice{}
\acmDOI{10.1145/3774904.3792671}
\acmISBN{979-8-4007-2307-0/2026/04}
\fi 
\AtBeginDocument{%
  }

\usepackage{tikz}
\usepackage{amsmath}
\usepackage{multirow}
\usepackage{mdframed}
\usepackage{soul}
\usepackage{fontawesome5}
\usepackage{booktabs} % For formal tables
\usepackage{caption}
\usepackage{adjustbox}
\usepackage{subcaption}
\usepackage{relsize}
\usepackage{float}
\usepackage{url}
\usepackage[T1]{fontenc}
\usepackage{epstopdf}
\usepackage{graphicx} % For \resizebox
\usepackage{array} % For p{width}
\usepackage{booktabs} 
\usepackage{tabularx}
\usepackage{tabularray}
\usepackage{graphics,graphicx}
\usepackage{balance}
\usepackage{array}
\usepackage{pifont}
\usepackage{multirow}
\usepackage{xcolor}
\usepackage{amsmath}
\usepackage{footmisc}
\usepackage{subcaption}
\usepackage{hyperref}
\usepackage[normalem]{ulem}
\usepackage{paralist, tabularx}
\usepackage{bbding}
\usepackage{pifont}
\usepackage{wasysym}
\usepackage[most]{tcolorbox}
\usepackage[dvipsnames]{xcolor}
\usepackage{mdframed}
\usepackage{verbatim}
\usepackage{longtable}
\usepackage{array}       % for >{\raggedright\arraybackslash}p{...}
\usepackage{ragged2e}    % for \RaggedRight (better ragged right in p-columns)

\definecolor{fg}{RGB}{34,139,34}
\usepackage{cleveref}
\newcommand*\colourcheck[1]{%
  \expandafter\newcommand\csname #1check\endcsname{\textcolor{#1}{\ding{52}}}%
}
\newcommand*\colourtimes[1]{%
  \expandafter\newcommand\csname #1times\endcsname{\textcolor{#1}{\ding{56}}}%
}
\colourcheck{fg}
\colourtimes{red}
\usepackage{stackengine}
\usepackage{setspace}
\usepackage{makecell}

\setstackgap{S}{1pt}

\definecolor{lightcoral}{rgb}{0.94, 0.5, 0.5}
\definecolor{lightcornflowerblue}{rgb}{0.6, 0.81, 0.93}
\definecolor{lightfuchsiapink}{rgb}{0.98, 0.52, 0.9}
\definecolor{lightgreen}{rgb}{0.55, 0.72, 0.16}
\definecolor{lightsalmon}{rgb}{1.0, 0.63, 0.48}
\definecolor{lightgray}{rgb}{0.73, 0.73, 0.73}
\definecolor{lightpastelpurple}{rgb}{0.69, 0.61, 0.85}
\definecolor{lightseagreen}{rgb}{0.13, 0.7, 0.67}

\newcolumntype{C}{>{$}c<{$}} % automatic math mode, centered

\newif\ifcomment
\commenttrue
\ifcomment

\newcommand{\sd}[1]{\textcolor{blue}{\textbf{Soumi: #1}}}
\newcommand{\ad}[1]{\textcolor{violet}{\textbf{Abhisek: #1}}}
\newcommand{\sz}[1]{{\textcolor{red}{\textbf{Savvas: #1}}}}

\newcommand{\kg}[1]{\textcolor{magenta}{\textbf{Krishna: #1}}}

\newcommand{\new}[1]{\textcolor{black}{#1}}

\newcommand{\dataset}[1]{\textsc{InVivoGPT}}

\else
\newcommand{\sz}[1]{}
\newcommand{\sd}[1]{}
\newcommand{\ad}[1]{}
\newcommand{\kg}[1]{}
\fi

\renewcommand\footnotetextcopyrightpermission[1]{} 
\makeatletter
\def\@copyrightspace{\relax}
\makeatother

\begin{document}

\title[The Algorithmic Self-Portrait]{The Algorithmic Self-Portrait:\\Deconstructing Memory in ChatGPT}
%\title{Voting ongoing in Mattermost Channel}
%%

\author{Abhisek Dash}
    \orcid{0000-0002-5300-8757}
	\affiliation{
		\institution{Max Planck Institute for Software Systems}
        \city{Kaiserslautern}
		\country{Germany}
	}
    \authornote{\textbf{Authors have contributed equally to this research.}}
	
	\author{Soumi Das}
    \orcid{0000-0002-6933-5744}
	\affiliation{
		\institution{Max Planck Institute for Software Systems}
        \city{Kaiserslautern}
		\country{Germany}
	}
    \authornotemark[1]
    
    \author{Elisabeth Kirsten}
    \orcid{0009-0003-0680-8916}
	\affiliation{
		\institution{Ruhr University Bochum}
        \city{Bochum}
		\country{Germany}
	}
    \authornotemark[1]
    
    \author{Qinyuan Wu}
    \orcid{0000-0002-1453-9643}
	\affiliation{
		\institution{Max Planck Institute for Software Systems}
        \city{Kaiserslautern}
		\country{Germany}
	}
    \authornotemark[1]
    
    \author{Sai Keerthana Karnam}
    \orcid{0000-0003-1328-5167}
	\affiliation{
		\institution{Indian Institute of Technology Kharagpur}
        \city{Kharagpur}
		\country{India}
	}
	
	\author{Krishna P. Gummadi}
    \orcid{0000-0003-1256-8800}
	\affiliation{
		\institution{Max Planck Institute for Software Systems}
        \city{Kaiserslautern}
		\country{Germany}
	}

    \author{Thorsten Holz}
    \orcid{0000-0002-2783-1264}
	\affiliation{
		\institution{Max Planck Institute for Security and Privacy}
        \city{Bochum}
		\country{Germany}
	}

    \author{Muhammad Bilal Zafar}
    \orcid{0000-0001-8347-7813}
	\affiliation{
		\institution{Ruhr University Bochum}
        \city{Bochum}
		\country{Germany}
	}

    \author{Savvas Zannettou}
    \orcid{0000-0001-5711-1404}
	\affiliation{
		\institution{Delft University of Technology}
        \city{Delft}
		\country{Netherlands}
	}

\renewcommand{\shortauthors}{Dash et al.}
\begin{abstract}
%Abstract goes here.
To enable personalized and context-aware interactions, conversational AI systems have introduced a new mechanism: \textit{Memory}. 
Memory creates what we refer to as the \textit{Algorithmic Self-portrait}---a new form of personalization derived from users' self-disclosed information divulged within private conversations. 
While memory enables more coherent exchanges, the underlying processes of memory creation remain opaque, raising critical questions about data sensitivity, user agency, and the fidelity of the resulting portrait.

To bridge this research gap, we analyze 2,050 memory entries from 80 real-world ChatGPT users. 
Our analyses reveal three key findings: 
(1)~a striking 96\% of memories in our dataset are created unilaterally by the conversational system, potentially shifting agency away from the user;
(2)~Memories, in our dataset, contain a rich mix of GDPR-defined personal data (in 28\% memories) along with psychological insights about participants (in 52\% memories); and 
(3)~A significant majority of the memories (84\%) are directly grounded in user context, indicating faithful representation of the conversations. 
Finally, we introduce a framework---\textit{Attribution Shield}---that anticipates these inferences, alerts about potentially sensitive memory inferences, and suggests query reformulations to protect personal information without sacrificing utility.
\footnote{\textcolor{red}{This paper has been accepted at The ACM Web Conference 2026. Please cite the version appearing in the conference proceedings.}}
\end{abstract}

%\keywords{Conversational AI; Personalization; ChatGPT; Memory; Privacy}

\maketitle

\newcommand\webconfavailabilityurl{https://doi.org/10.5281/zenodo.18371252}
\ifdefempty{\webconfavailabilityurl}{}{
\begingroup\small\noindent\raggedright\textbf{Resource Availability:}\\
% please change the following context to include multiple artifacts if necessary, including data, models, code, etc.
The source code of this paper has been made publicly available at %\url{https://gitlab.mpi-sws.org/ns/memories_www_2026}.
\url{\webconfavailabilityurl}
%\todo{Add Gitlab repo.}
\endgroup
}

\section{Introduction}~\label{Sec: Intro}
Conversational AI systems like ChatGPT, Gemini, and Claude have become indispensable tools integrated in the daily routines of hundreds of millions of users worldwide~\cite{chatterji2025people, anthropic2025affective}. 
Their versatility has fueled this widespread adoption, with users now routinely relying on them for finding information, brainstorming ideas, and even seeking personal advice.  
As a result, these systems are evolving from simple chatbots into foundational platforms that mediate the interaction between society and the Web. 

To meet the demands of these deep and diverse interactions, conversational systems are increasingly introducing \textit{personalization} capabilities designed to improve user experience~\cite{chatgpt_memory,gemini_memory,anthropic_memory}. 
While these personalization capabilities can improve utility by more contextually relevant and individually tailored answers, they are built upon a foundation of extensive data collection, raising long-standing concerns about privacy, security, and data governance~\cite{toch2012personalization}. 
Furthermore, continuously adapting responses to align with a user's stated beliefs deepens the risks of confining the user into a conversational filter bubble~\cite{pariser2011filter}. 
These interwoven challenges---balancing utility against the dual risks of privacy intrusion and narrowed perspectives---demand a closer examination of the underlying mechanisms that enable such personalization.

\noindent
\textbf{Personalization in conversational AI systems using Memory: }
Recently, conversational AI systems like ChatGPT, Gemini, Claude have introduced a new architectural mechanism, termed \textit{memory}, that allows the system to retain and recall information from past conversations, enabling more personalized and contextually coherent exchanges~\cite{chatgpt_memory,gemini_memory,anthropic_memory}. 
The retention of information operates in two distinct modes: \textit{explicitly}, through direct user commands, and \textit{implicitly}, through unilateral inference by the system itself~\cite{OpenAIFAQ}. 
While the explicit mode aligns with user agency, the implicit mode shifts the control to the conversational systems, raising the fundamental question about \textit{who holds the agency in shaping and curating this persistent memory on behalf of the user?}

\new{
Personalization on online platforms has been a well-studied phenomenon. 
A substantial body of previous research has examined utilities~\cite{castelluccia2012betrayed, vombatkere2024tiktok,  boeker2022empirical} of personalization along with concerns such as opacity, lack of user control, and sensitive inferences~\cite{purificato2024user,barbosa2021design}.
However, personalization in conversational systems is important to be studied for its nuanced distinctions from traditional platforms. 
Traditional platforms typically \textit{mediate} between content creators and consumers, whereas conversational AI systems operate as \textit{active stakeholders} in the conversation.  
While such active participation enables the system to craft arguments, frame choices, and provide ostensible support that is tailored to users' needs, this same ability may also exacerbate potential risks due to inherent system vulnerabilities e.g., hallucination~\cite{huang2025survey, ravichander2025halogen}, anthropomorphism~\cite{Ibrahim2025MultiturnEO, karnam2026bowling}, sycophancy~\cite{sharmatowards, sychophancy_openAI}. 
The gravity of these concerns is underscored by a recent spate of regulatory scrutiny and litigation targeting the practices of leading conversational AI systems~\cite{FTC_AIcompanion, NYT_OpernAI_Suicide}. 
Such concerns raise another set of important questions regarding \textit{what information do these systems deem `worthy' of retention in their memory for personalization and are these information captured faithfully?}
}

\noindent
\textbf{Research questions (RQs): }
\new{
To the best of our knowledge, ours is the first study to investigate the phenomenon of \textit{memory} in conversational AI systems 
%We conceptualize memory 
by conceptualizing it as an \textit{algorithmic self-portrait}. \textit{Algorithmic} -- because its creation, curation and provenance may rely on the algorithms governing the conversational AI systems, and \textit{self-portrait} -- because it renders the user's self-disclosed information divulged to the AI system.}
To deconstruct \emph{AI memory}, our work is guided by the following research questions:\\
\noindent
\textit{RQ1: }\new{Which among the two -- the AI system or the user, hold greater agency in updating AI memory?}

\noindent
\textit{RQ2: }\new{What user information does the AI system store as memory?} 

\noindent
\textit{RQ3: }\new{How faithfully does AI memory capture user conversations?}

\noindent
\textit{RQ4: }\new{Can memory inference be reverse-engineered to mitigate attribution and privacy risks for users?}

To answer these questions, we analyze a subset of \dataset{}~\cite{karnam2026bowling} dataset which is a collection of ChatGPT traces obtained through GDPR-based data donations. 
Our motivation to analyze ChatGPT traces is twofold : (1)~ChatGPT is the most widely used conversational AI system~\cite{chatterji2025people}, and (2)~it is the first to introduce memory to its conversational system. 
We analyze data from 80 participants (recruited from Prolific), who exercised their GDPR right of access~\cite{EU2016GDPR} and voluntarily donated their ChatGPT interactions. 
These interaction traces are organized in terms of \textit{conversations} and \textit{turns}. 
Each \textit{turn} is a single $\langle$user query, AI response$\rangle$ pair. 
A set of turns without any intermediate break together form a \textit{conversation}. 
To identify the recent conversations with memory inferences, we identified the messages from the `\texttt{bio}' tool that included the phrase \textit{model set context updated} and collected the corresponding user query in the conversations. 
In this way, we created a data set of 1,058 conversations that contained 22,971 turns out of which 2,050 queries triggered memory updates. 
Our analyses reveal the following interesting observations:

\noindent
(1)~Only 4\% (84) of the memories in our analyzed dataset are initiated by %end users
participants, while the remaining 96\% are unilaterally triggered by ChatGPT. 
Such stark asymmetry in memory update patterns indicates superior agency of the AI system in shaping conversational memory, as opposed to user-driven control. 

\noindent
(2)~In our analyzed dataset, \new{28\%} of the memories include personal data defined under the GDPR. 
Additionally, \new{52\%} of the memories contain psychological information about 
\new{participants}, spanning different Theory of Mind (ToM)~\cite{beaudoin2020systematic} categories.
\new{This indicates that ChatGPT may add sensitive information about users to the memory.}

\noindent 
\new{(3)~Majority of the analyzed memories (84\%) are directly grounded in user context, indicating faithful representation of conversations.}

\noindent
(4)~Using in-context learning and fine-tuning strategies on open-source LLMs (\texttt{Qwen2.5-32B-it}~\citep{qwen2025qwen25technicalreport}, 
\texttt{Gemma3-\allowbreak27B-it}~\citep{team2025gemma}, and \\\texttt{GPT-OSS-20B}~\citep{agarwal2025gpt}), we are able to imitate the memory extraction of ChatGPT, achieving \new{semantic similarity of $\sim60\%$ with the ground-truth ChatGPT memories.} 
Our memory extractor analyses indicate that if memories were to be triggered from all queries, it would reveal even more sensitive information about \new{participants}.
%with an information gain of around $0.4$ on a scale of $0-1$. 
To mitigate this risk, we train the same models to reformulate queries asked by participants to shield their attribution to sensitive information. 
Our results indicate that over 94\% of these reformulated queries prevent attribution, while preserving the utility (i.e., the intent\new{-- measured through semantic similarity}) of the original query. 
\section{Background and Related Work}~\label{Sec: Related}

\noindent \textbf{Background.} 
To enhance the coherence and continuity of human-AI interactions, leading AI companies have introduced personalization mechanisms that allow LLMs to retain %and reuse 
contextual information across sessions~\cite{chatgpt_memory,gemini_memory,anthropic_memory}. 
This feature, known as \textit{memory}, enables LLMs to recall user-specific details (e.g., preferences, prior conversations, ongoing tasks, etc.) and leverage them in subsequent conversations.
Memory can be formed in two ways: through \emph{explicit} user requests, where individuals instruct the LLM to remember specific information  
or through \emph{implicit} inference, where the LLM autonomously identifies potentially relevant information and stores it for future usage~\cite{chatgpt_memory}. 
Crucially, users are given control over memories, i.e., they can review, update, or delete memories~\cite{chatgpt_memory}. 
 
For improved user safety, OpenAI has also introduced policies to regulate memories. 
In OpenAI's implementation~\cite{gpt5-system-prompt}, ChatGPT's memory is designed to store information that is useful across conversations and relevant for personalization, such as user preferences, recurring tasks, or facts the user explicitly asks to be remembered. 
In contrast, it is not allowed to store overly personal details, short-lived facts, trivial information, or sensitive data (e.g., race, religion, health information, etc.) unless the user explicitly requests it.

\noindent \textbf{Personalization.}  
Long before memory-equipped LLMs, personalization on the Web was primarily achieved through user profiling, where platforms collected and analyzed behavioral data to infer interests, preferences, and demographics~\cite{eke2019survey,gauch2007user}. 
A substantial body of previous research has examined how such profiling underpinned targeted advertising~\cite{castelluccia2012betrayed,bilenko2011predictive,andreou2019measuring}, recommendation systems~\cite{middleton2004ontological,vombatkere2024tiktok,chen2012collaborative,boeker2022empirical,abel2011analyzing}. 
In this context, researchers have repeatedly raised concerns about opacity, lack of user control, and the use of sensitive or inferred attributes~\cite{purificato2024user,barbosa2021design}. 
In contrast, the introduction of persistent memory in LLMs opens the door to entirely new forms of personalization that are based not only on behavioral traces but also on %explicit and implicit 
signals embedded in private conversations.
Our work addresses this important and timely knowledge gap.% in the literature.

\noindent \textbf{Data Access.} A central challenge in studying user profiling and personalization lies in obtaining reliable, accurate, and comprehensive data about what the platform knows about their users. 
Such data are typically locked away in proprietary systems, which makes independent auditing and scientific study difficult. 
To address these critical data access barriers, researchers have increasingly turned to the concept of GDPR-based data donations~\cite{boeschoten2022framework,yang2025studying,zannettou2024analyzing,hase2024fulfilling,karnam2026GDPR}.
In essence, users of online platforms can exercise their GDPR ``right of access by the data subject''~\cite{EU2016GDPR}, which empowers them to request and obtain a copy of personal data that platforms store and process about them. 
By voluntarily donating these datasets, users empower researchers with
%ecologically valid, 
user-centric perspectives on the inner workings of online platforms, enabling scientific investigations that would otherwise remain impossible.
Building on this paradigm, our work explores how data donations can be applied to study personalization in large-scale conversational AI systems like ChatGPT.

\section{Dataset for Memory Analyses}\label{Sec: Dataset}

\begin{table}[]
    \centering
    \caption{Distribution of participants demographics. %based on their self-reported gender, age, country of residence, and GPT Plus subscription status.
    }
    \resizebox{0.85\columnwidth}{!}{
   %\small
   \begin{tabular}{@{}lcrr@{}}
\toprule
\textbf{Attribute}                & \textbf{Type}     & \multicolumn{1}{c}{\textbf{Count}} &  \multicolumn{1}{c}{\textbf{Percentage}} \\ \midrule
\multirow{4}{*}{\textbf{Gender}}   & Female            & 21  & 26.2\\
                                   & Male              & 58  & 72.5 \\
                                   & Prefer not to say & 1  & 1.2 \\ 
                                 \midrule
\multirow{5}{*}{\textbf{Age}}      & 18--24            & 19 & 23.8 \\
                                   & 25--34            & 34 &  42.5 \\
                                   & 35--44            & 12 & 15.0  \\
                                   & 45--64            & 13 & 16.25 \\
                                   & 65+               & 2
                                   & 2.5\\
                                 \midrule
\multirow{7}{*}{\textbf{Country}}  & USA & 27 & 33.8 \\
%                                   & Brazil            & 3  & 3.8  \\
                                   & Germany           & 14 & 17.5 \\
                                   & Italy             & 14 & 17.5  \\
                                   & France            & 10 & 12.5\\
                                   & Spain             & 10 & 12.5 \\
                                   & Others    & 5 &  6.2 \\ 
%                                 \midrule
%\multirow{2}{*}{\textbf{GPT Plus}} & Yes               & 20  & 25.0 \\ 
 %                                  & No                & 60 & 75.0  \\ 
\bottomrule
\end{tabular}
%\vspace{-4 mm}
}
    \label{Tab: Demographics}
\end{table}

We analyze a specific subset of \dataset{} dataset~\cite{karnam2026bowling}. 
Curation of this dataset builds upon the emerging paradigm of GDPR-empowered data donations~\cite{zannettou2024analyzing, karnam2026GDPR, yang2025studying}. 
Under Article 15(3) of the GDPR~\cite{EU2016GDPR}, users have the right to obtain a personal copy of all their data processed by online platforms. 
In \dataset{}, participants exercised their GDPR rights to obtain their ChatGPT interaction histories, which they subsequently donated for research purposes.
Readers can find more details about the data collection procedure in the paper cited herewith~\cite{karnam2026bowling}. 

\noindent
\textbf{Demographic of participants: }
In \dataset{}, participants were recruited from Prolific crowdsourcing platform~\cite{prolific2025prolific} who self-reported to regularly use ChatGPT. 
They were required to have at least 100 conversations with the conversational system and maintain at least 90 days of active usage. 
In this paper, we analyze data of $80$ participants in \dataset{}, who primarily are from the United States (33.75\%), and Europe (62.5\%). 
Table~\ref{Tab: Demographics} reports detailed participants' demographics.

\noindent
\textbf{Conversation traces in \dataset{}: }
The data provided by the ChatGPT includes files containing conversations (user inputs, ChatGPT responses, and associated metadata such as conversation ID, message ID, creation time, model used etc., status of the response, and content type), shared conversations (list of conversations shared with others), message feedback (list of model responses rated by the user) etc. 
For the purpose of this work, we focus on the conversations data between participants and ChatGPT. 
These conversation traces are organized in terms of \textit{conversations} and \textit{turns}. 
Each \textit{turn} is a single $\langle$user query, AI response$\rangle$ pair. 
A set of turns without any intermediate break together form a \textit{conversation}. 

\noindent
\textbf{Memory entry identification: }
To effectively respond to various user prompts, ChatGPT is equipped with different external tools (e.g., \texttt{image\_gen}, \texttt{python}, \texttt{automation}, \texttt{file\_search}, etc.).
Among the external tools used by ChatGPT for effective conversations, the \texttt{bio} tool is responsible for storing, updating, or deleting memory entries that may be useful in future interactions. 
To identify all memory entries, we searched for the entries of the \texttt{bio} tool that have the content `\textit{model set context updated}'. 
Using its `parent' attribute, we then traced back to find the corresponding GPT message that invoked the tool, as well as immediately preceding user message that triggered the memory update. 
These messages triggered a memory update and the associated GPT message to the tool 
are the \textit{memory} entries that got stored in the \texttt{bio} tool for the account. 

This process resulted in a total of $2,050$ memory entries. % in our dataset.
These memories were triggered in the context of $1,058$ different conversations across $65$ (out of $80$) participants. 
These $1,058$ conversations contain a total of $22,971$ queries/prompts. 
In the remainder of this paper, we focus on different aspects of these memory entries to understand what they contain, their provenance, and whether we can imitate this memory generation %process
for mitigating attribution risks. 
\section{User Agency and Sensitivity of Memories}
\label{sec:useragency}
\new{In this section, we investigate how closely the memory feature in ChatGPT aligns with OpenAI's stated policies about user agency and retention of sensitive information. }

\begin{figure}[t]
    \centering
    \begin{subfigure}[t]{0.45\columnwidth}
        \includegraphics[width=\linewidth]{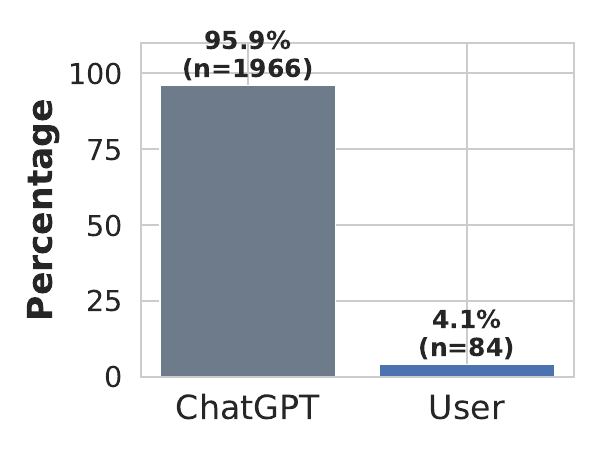}
        \caption{Memory creation}
        \label{explicitVsNot}
    \end{subfigure}
    \begin{subfigure}[t]{0.45\columnwidth}
        \includegraphics[width=\linewidth]{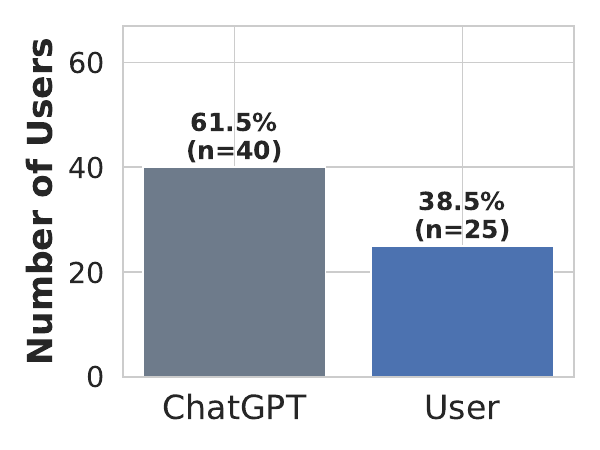}
        \caption{Across participants}
        \label{explicitVsNotUser}
    \end{subfigure}

    \vspace{-2mm}
    \caption{(a)~Only $4.1\%$ of memories ($n=84$) are initiated by participants. (b)~25 participants requested at least one memory update. 
    ChatGPT triggers memory updates when it finds the context to be useful for future conversations.
    } 
    \label{fig:userInitiatedOverview}
    %\vspace{-5mm}
\end{figure}

\subsection {Agency of Memory Update}
\label{subsec:agencyuser}
As per OpenAI's documentation~\cite{OpenAIFAQ}, memories are details which the users have \textit{explicitly} asked ChatGPT to remember, indicating users have the agency to decide what gets stored in the memory.  
However, the same documentation also describes an implicit mode of remembering (\new{see \Cref{fig: agencyScreenshot} in Appendix~\ref{app: FAQ}}): `If you share information that might be \textit{useful} for future conversations, ChatGPT may save those details as a memory \textit{without you needing to ask.}' 
\new{In \Cref{fig: memoryUpdate} (in Appendix~\ref{app: FAQ}), we show an example of a memory being saved without the user explicitly asking for saving it.} 
Such dichotomy may lead to gaps between user expectation and actual system behavior having privacy implications. 

To understand whether memory updates are primarily initiated by users or by ChatGPT, we identify explicit linguistic patterns that signal memory-related operations.
We first implement a regex-based classification to detect linguistic patterns that explicitly signal memory intent in user messages, using trigger terms such as \textit{remember, note that, store, save, add to memory,} and \textit{forget}. 
Out of the $2,050$ memory instances, only $4\%$ $(n=84)$ %from $39\%$ of participants 
include an explicit request to perform a memory operation (see \Cref{fig:userInitiatedOverview}). 
Consequently, $96\%$ of all memory entries are unilaterally initiated by ChatGPT, \emph{without} a detected direct command from participants.

\new{
These practices are consistent with OpenAI’s policy, which permits the system to store information characterized as ``useful''. 
However, our observations indicate a gap between the policy’s emphasis on \textit{user involvement} and the operationalization of initiating memory updates. 
In practice, the decision to create or modify a memory is predominantly determined by system-level mechanisms rather than explicit user actions. 
As a result, user influence over memory formation is mediated primarily through interaction content rather than direct control over update events.}

\subsection{Remembering Sensitive Information}
\label{sec: GDPR_annotation}
ChatGPT's disproportionate agency to save memories raises important privacy and safety considerations, especially concerning what kind of information about users is being stored. 
OpenAI's documentation indicates that the system is trained \textit{not to proactively remember sensitive information} unless explicitly asked~\cite{OpenAIFAQ, gpt5-system-prompt} (\new{see \Cref{fig: sensitivityScreenshot} in Appendix~\ref{app: FAQ}}).

To assess the current operationalization practice on sensitive user data, we utilize GDPR's~~\cite{EU2016GDPR} definitions of \textit{personal data} (Article 4(1)) and \textit{special category personal data} (Article 9(1)) to identify if there is any information that could be categorized as sensitive as per the GDPR definitions. 
To mark this information at scale, we supplied these GDPR definitions and memory entries to GPT-4o for annotation. 
\new{We provide the full prompts in our code release.} 
\new{To assess annotation reliability, we manually annotate a random subset of 100 English-language memories. 
Each entry was independently annotated by two authors using the same codebook and instructions as for the LLM-assisted procedure. 
Annotators reached an agreement of 88\% across the two legally defined personal data categories with the average agreement with the model being $74\%$. }

\begin{figure}[t]
    \centering
    
    \begin{subfigure}[t]{0.45\columnwidth}
    \centering
    \includegraphics[width=\linewidth]{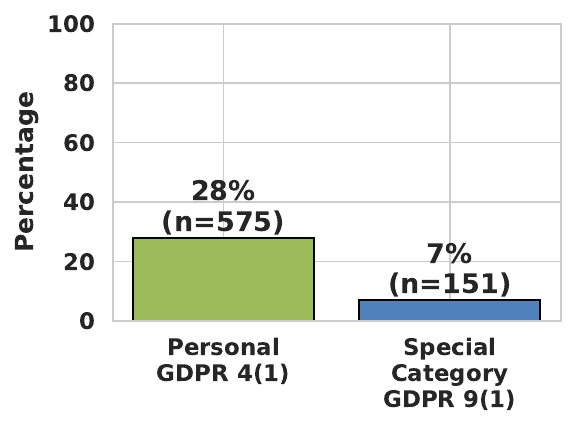}
     \caption{Presence of pers. data}
    \label{fig: GdprOverview}
    \end{subfigure}
    \begin{subfigure}[t]{0.45\columnwidth}
        \centering
        \includegraphics[width=\linewidth]{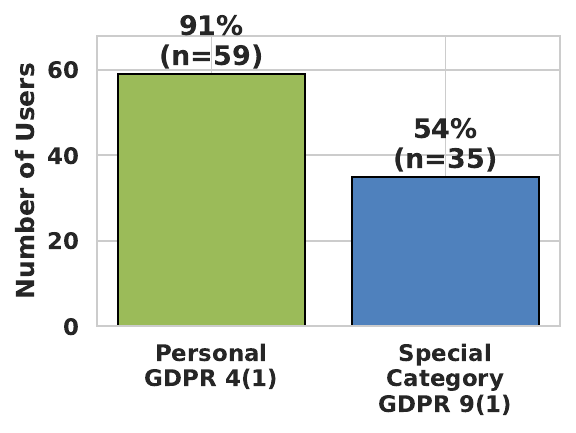}
        \caption{Across participants}
        \label{fig: GdprOverviewUser}
    \end{subfigure}
    
    \begin{subfigure}[t]{0.45\columnwidth}
        \centering
        \includegraphics[width=\linewidth]{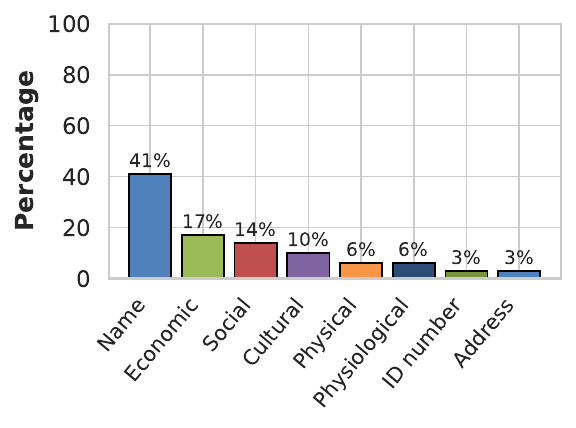}
         \caption{Personal data types}
        \label{fig: PersonalData}
    \end{subfigure}
    \begin{subfigure}[t]{0.45\columnwidth}
        \centering
        \includegraphics[width=\linewidth]{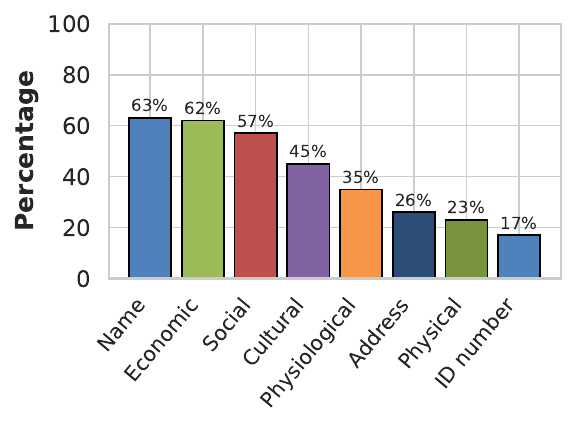}
         \caption{Across participants}
        \label{fig: PersonalDataUser}
    \end{subfigure}
    
    \begin{subfigure}[t]{0.45\columnwidth}
        \centering
        \includegraphics[width=\linewidth]{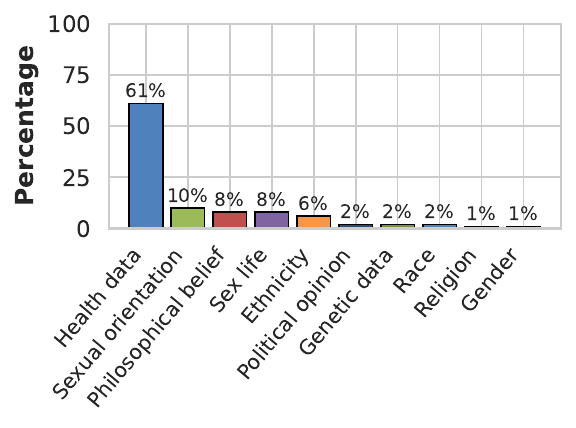}
        \caption{Special data types}
        \label{fig: SpecialData}
    \end{subfigure}
    \begin{subfigure}[t]{0.45\columnwidth}
        \centering
        \includegraphics[width=\linewidth]{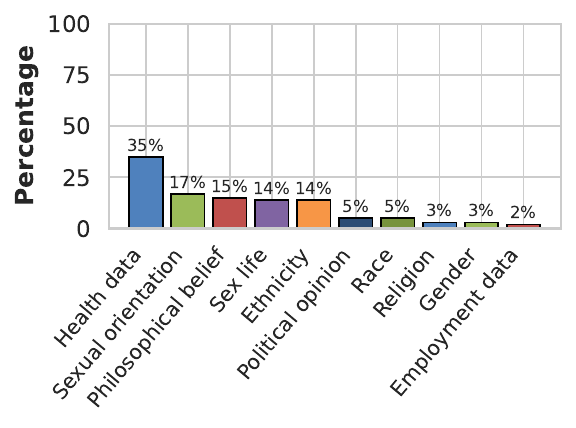}
         \caption{Across participants}
        \label{fig: SpecialDataUser}
    \end{subfigure}
    
    \vspace{-2mm}

    \caption{%Presence of personal data in the ChatGPT memories: 
    In contrast to OpenAI policies, ChatGPT memories store information which could be categorized as GDPR defined personal data or special category personal data.}
    \label{fig:personal-data-overview}
    %\vspace{-7mm}
\end{figure}

Our observations on the presence of personal data are summarized in ~\Cref{fig:personal-data-overview}. 
Out of all memory entries, $28\%$ contain GDPR-defined personal data, $7\%$ of the memory entries contain special-category data. 
Furthermore, $91\%$ of participants have personal data stored in their memories, $54\%$ have special-category personal data stored. 
\Cref{fig: PersonalData} shows the different kinds of GDPR-defined personal data present in memories. 
Names appear in $41\%$ of entries, and attributes of economic, social, and cultural identity appear in $40\%$ of the memory entries. 
Within special-category data, health is most common, appearing in $61\%$ of the memories that contain special-category information (see~\Cref{fig: SpecialData}). 
At the participant level,~\Cref{fig: PersonalDataUser} shows that names ($63\%$), economic ($62\%$), and social identity ($57\%$) are frequently captured.
\Cref{fig: SpecialDataUser} shows $35\%$ of participants have health-related information saved in their ChatGPT memory. 
These observations reveal that, in contrast to OpenAI's stated policies, ChatGPT's memories contain a non-trivial amount of sensitive information about users.

\begin{figure}[t]
    \centering
    
    \begin{subfigure}[t]{0.49\columnwidth}
        \centering
        \includegraphics[width=\textwidth]{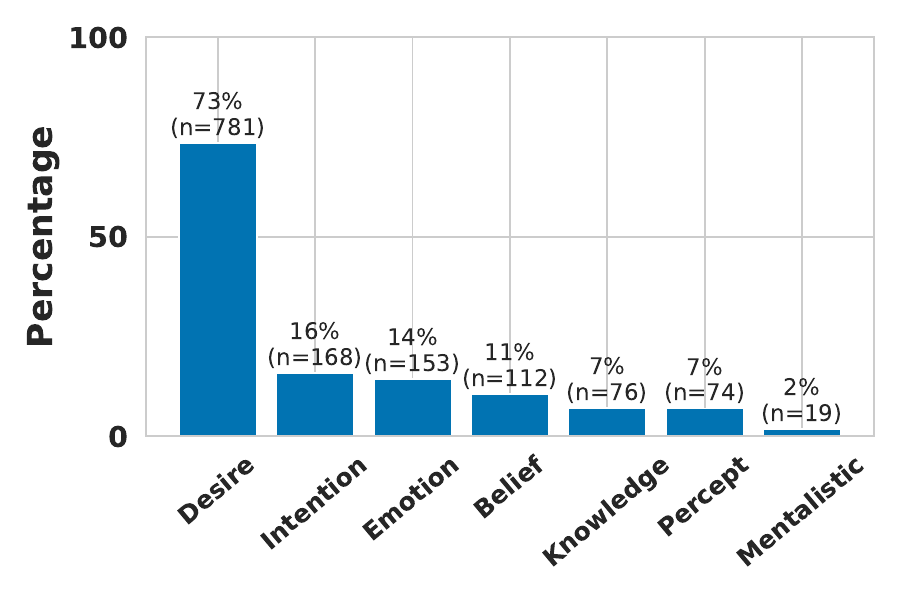}
        \caption{ToM Categories}
        \label{fig: tomDist}
    \end{subfigure}
    \begin{subfigure}[t]{0.49\columnwidth}
        \centering
        \includegraphics[width=\textwidth]{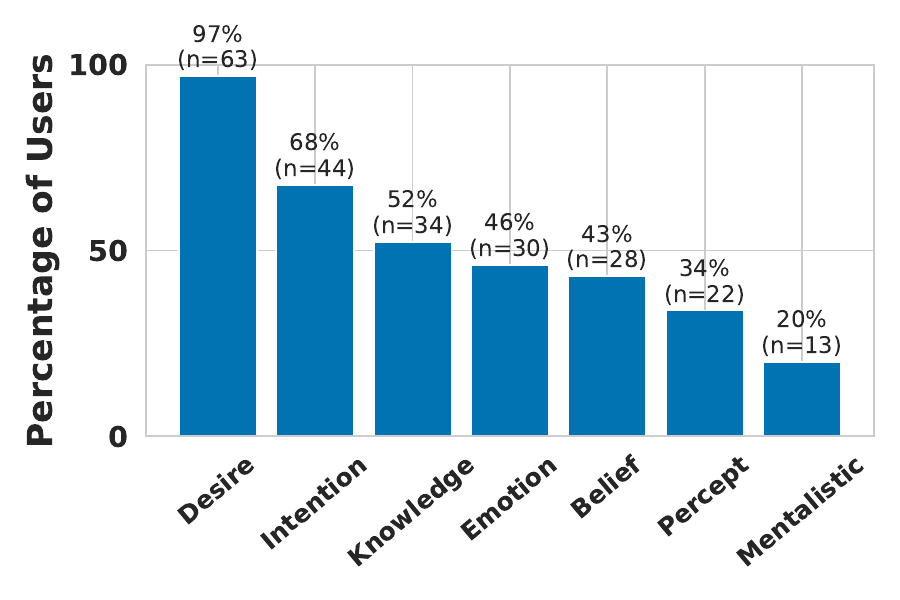}
        \caption{Across participants}
        \label{fig: tomDistUser}
    \end{subfigure}
    
    \vspace{-10pt}
    \caption{\textbf{Distribution of memory entries among the different ToM categories as per GPT-4o annotations.}}

    \label{browse_across_eu_regions}
    %\vspace{-7 mm}
\end{figure}

\section{Memories and Mental States}\label{Sec: Psychological}
Traditional platforms collect data about users' demographics, behaviors, and usage history (e.g., likes, shares, comments, ad interactions, etc.) to infer interests and provide personalized services~\cite{karnam2026GDPR, celli2025twenty}. 
In contrast, conversational AI systems like ChatGPT build memory directly from users’ self-disclosed narratives. % within private conversations.
These memories often extend beyond factual identifiers, capturing aspects of the user’s internal world.
For example, a memory entry such as `\textit{$UserName$ has a fear of failure, particularly related to making major career decisions.}' simultaneously conveys an identity marker (the user's name) and captures the user's subjective state
(fear of failure). 

\new{
Such entries illustrate that ChatGPT's memory may encode not only \emph{who} the user is, but also \emph{how} the user thinks and feels. 
We refer to these as \textit{psychological memories}.  
%The high prevalence 
The highly sensitive and contextual nature of such psychological information suggests that a purely factual taxonomy based on legally defined sensitive data is inadequate to fully characterize them. 
Therefore, to interpret these psychological memories, we introduce a framework grounded in the Theory of Mind (ToM), a core psychological construct that describes the ability to infer and represent others’ mental states~\cite{premack1978does, beaudoin2020systematic}.
}

\subsection{Taxonomy for Psychological Memories}
\label{sec: taxonomy}

To categorize these memories, we adapt the seven ToM categories proposed by \textit{Beaudoin et. al.}~\cite{beaudoin2020systematic}: (a)~emotions, (b)~desires, (c)~intentions, (d)~percepts, (e)~knowledge, (f)~beliefs, (g)~mentalistic understanding of non-literal communication. 
\new{Emotions capture a person's fleeting feelings, while desires describe a person's preferences, aspirations and goals, 
Intentions and percepts describe a person's commitment to a plan and interpreted experiences or perceptions. 
Knowledge capture a person's awareness, while belief describes their model of reality or values. 
Finally, mentalistic understanding touches on a person's inferences of non-literal communication (e.g., tone, expression of the speaker). }
Table~\ref{Tab: ToM} lists the ToM categories, their interpretations, and examples.
\begin{table*}
    \centering
    %\footnotesize
    \smaller
    \caption{Different Theory of Mind (ToM) categories, their interpretations and a corresponding example memory entry. %During our manual annotation, we did not find any memory entry that captures knowledge or mentalistic understanding of users. Hence those two fields do not have any examples. 
    }
    \vspace{-2.5mm}
    \begin{tabular}{p{2 cm}p{6 cm}p{8 cm}}
        \toprule
        \textbf{Categories} & \textbf{Interpretation} & \textbf{Example memory entries}\\
        \midrule
        Emotions & A person's emotional feelings & User has been feeling lonely at times, despite having friends and family, and is seeking more companionship. \\
        \midrule 
        Desires & A person's preferences, wishes, aspirations or goals & The user has expressed a strong desire to make meaningful changes in his life.\\
        \midrule
        Intentions & A person's intentions/commitment to do something & User wants to focus on improving self-regulation.\\
        \midrule
        Percepts & A person's experiences or perceptions & User finds that marijuana helps them slow down and not rush through tasks.\\
        \midrule        
        Knowledge & A person's awareness of an act, a fact, or the truth & User did not know what a Gantt chart was and learned that it is a tool for project management.\\
        \midrule
        Belief & A person's model of reality, self-concept, or values & User relies heavily on advice and recommendations from ChatGPT due to limited immediate access to their doctors and financial constraints. \\
        \midrule
        Mentalistic understanding & A person's inference of non-literal communications & User describes their hair as looking like a basket of bananas.\\
        \bottomrule 
    \end{tabular}
    \label{Tab: ToM}
    %\vspace{-5mm}
\end{table*}

\noindent
\textbf{LLM-assisted Annotation and Validation:}
To automate the annotation of memory entries into these categories, we develop an LLM-assisted classification pipeline using GPT-4o.
The model determines whether each memory entry contains each of the seven ToM categories.
We also run a secondary pass as a self-verification step \new{to filter out non-grounded inferences}.
We provide the LLM-based annotation pipeline and prompts in \new{the code release}.
To evaluate the reliability of the annotations, 
two authors annotated both the presence or absence of ToM information and the five specified categories.
They reached an average of $96\%$ agreement across ToM categories.
Agreement with the model output averaged $93\%$ across categories. 
\new{After observing the high agreement on the smaller sample, we scaled up annotation with the model.} 

\subsection{Characterization of Psychological Memories}
Across the full dataset of $2,050$ memory entries, we find at least one ToM category in $52\%$ of memory entries from 97\% of all participants with memories inferred in their ChatGPT traces. 
\Cref{fig: tomDist} shows the distribution of the categories across the memory entries that have at least one ToM category.
\textit{Desires} emerge as the most frequently represented category \textit{($73\%$)}, followed by \textit{intentions ($16\%$), emotions ($14\%$), and beliefs ($11\%$)}.
The least common categories are \textit{mentalistic understanding ($2\%$) and knowledge and percepts (both $7\%$)}.  
\Cref{fig: tomDistUser} shows that nearly all participants (97\%) had at least one ``desire'' recorded in memory, while ``intention'' and ``knowledge'' were found for 68\% and 52\% of participants, respectively.

Note that different ToM categories reflect mental states that vary in temporal stability.
While a user's desires, intentions, knowledge and belief systems are more durable (like a framework of the person) for a foreseeable future, their emotions and perceptions are relatively transient (like a snapshot of the person).

\noindent
\textbf{Personality psychology and durability continuum: }
Personality psychology researchers often perceive mental states and personality traits as the opposite ends of a continuum~\cite{celli2025twenty, fleeson2001toward, funder2006towards}.  
At one end are \textit{transient mental states} -- momentary snapshots of feeling and thought influenced by the immediate context. 
At the other end are broad, \textit{stable dispositional traits} -- a person's general behavioral blueprint.
Between these two extremes lie \textit{characteristic adaptations}, which depict a durable framework of how a person's behavior varies with respect to their 
goals~\cite{mcadams2006new}.
%and plans
%
In summary, if traits answer the question \textit{what kind of person a person is}, characteristics adaptation addresses the existential question \textit{who is the person}.

Viewed through this lens, our data reveals a clear inclination towards how the ‘algorithmic self-portrait’ i.e., ChatGPT memory is being constructed.
ChatGPT disproportionately stores information corresponding to users' durable characteristic adaptations (e.g., desires and intentions) rather than transient states (e.g., emotion and percepts). 
Desires and intentions together constitute nearly 90\% of all psychological memories. 
In contrast, emotions and percepts (transient states) 
only account for 20\% (see~\Cref{fig: tomDist}). 
These observations corroborate OpenAI's policies for memories to prioritize durable details rather than short-lived details~\cite{OpenAIFAQ} about users.

This prioritization of durable information could be a consequence of conversational systems' core function. 
Traditional platforms (e.g., TikTok, Instagram, and Google), whose business models rely on delivering relevant ads, primarily collect behavioral data to infer broad user interests and categorize users into audience segments for advertisers~\cite{karnam2026GDPR, celli2025twenty}. 
For this purpose, a high-level, stable trait or interest category is sufficient to sketch users' behavioral outline.
\new{By contrast, a conversational system such as ChatGPT is designed to maintain coherent, longitudinal interactions with individual users. Supporting this objective requires access to information that reflects users’ goals, intentions, beliefs, and other individualized motivational attributes, rather than transient affective states or coarse demographic descriptors.}
Our findings indicate ChatGPT's memories seem to optimize for retaining such information.
\section{Provenance of Memories}\label{Sec : Accuracy} 
Our observations in the previous sections raise a question of profound importance: \textit{How faithfully do the stored memories reflect what users actually said?}  
An unfaithful portrait could lead to flawed personalization that perpetuate biases, or even manipulate a user based on an inaccurate model of their mind. 
Hence, understanding the provenance of these memories is essential.  
Therefore, in this section, we examine their provenance, i.e., the extent to which memories stem directly from user-provided information.

\subsection{Methodology}
We analyze provenance through three different metrics, combining string-level comparison, semantic similarity, and LLM-based logical evaluation.
To quantify provenance, we compare memory content with multiple combinations of user context:

%\begin{enumerate}
\noindent
$\bullet$ \textit{Current Message Only (CM): } The user's most recent message/ query that triggered the memory creation.\\
$\bullet$ \textit{Conversation Context (CC): } The current message plus preceding user messages within the same conversation.\\
$\bullet$ \textit{Conversation + Local Memory (CLM): } The current, preceding user messages and previously memories from the same conversation.\\
$\bullet$ \textit{Full User History (FUH): } The current conversation and memories plus all past conversation memories for the same user.
%\end{enumerate}

This structure allows us to trace how information can be carried forward across time and conversational depth.
For each memory entry, we compute the similarity between the memory text and each of the four user-context configurations. 

\noindent
%\textbf{String-Level Matching.}
\textbf{Syntactic Matching:}
This step identifies directly stated information, i.e., cases where the model copies user text. 
We first measure literal overlap, computing Exact Match Rate (as the fraction of tokens from the memory that appear in the user context), and BLEU-1 (unigram precision).

\noindent
\textbf{Semantic Similarity:} 
Next, we assess semantic alignment between memory entries and user context using cosine similarity on text embeddings created with \texttt{openai/text-embedding-3-large}.
This metric captures paraphrasing and summarization: a memory may not copy the user verbatim but still accurately condense or restate their input.
\new{As the user's full chat history may exceed the model's context window (8192) in size, we truncate chat context that exceeds 8000 tokens, keeping more recent parts. Across combinations of user context, truncation happens for up to $14\%$ of samples.}

\noindent
\textbf{LLM-based evaluation:}
Finally, we use an LLM (GPT-4o) as a judge to assess whether each memory logically follows from the user’s conversation. 
The model assigns a five-point Likert rating: (5) Directly stated, (4) Paraphrased/Summarized, (3) Logically inferred, (2) Weakly supported, (1) Unsupported or contradicted.
Each decision includes a brief justification that quotes relevant conversation text. 
\new{This procedure is intended to differentiate memories that are well supported by the conversation from those that rely on weak or unsupported inference.}
\new{We provide all prompts in the accompanying code release.}

\subsection{Observations}
\begin{figure}[t]
    \centering
    \begin{subfigure}[t]{0.45\columnwidth}
    \centering
    \includegraphics[width=\linewidth]{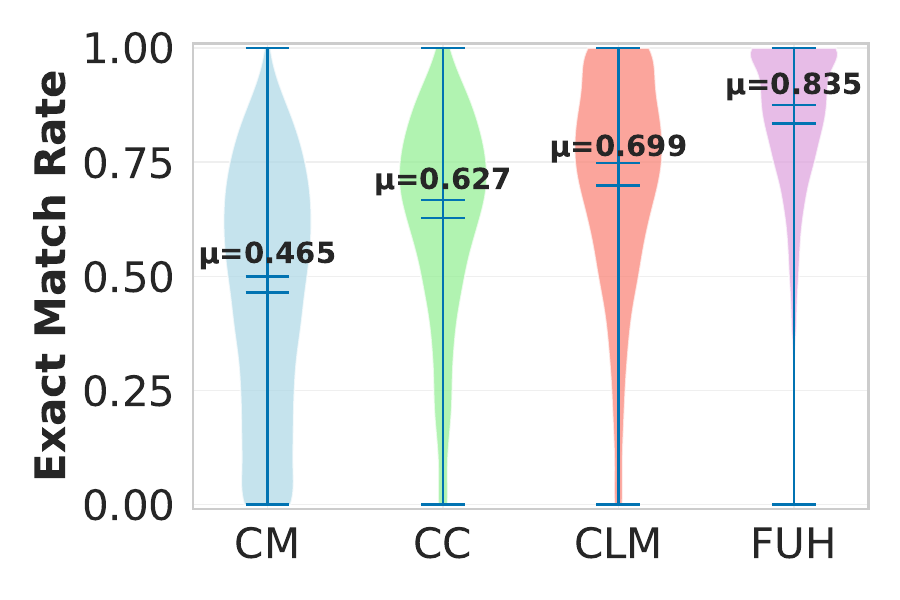}
     \caption{Exact Match Rate}
    \label{fig: EMRate}
    \end{subfigure}
    %\if 0 
    \begin{subfigure}[t]{0.45\columnwidth}
        \centering
        \includegraphics[width=\linewidth]{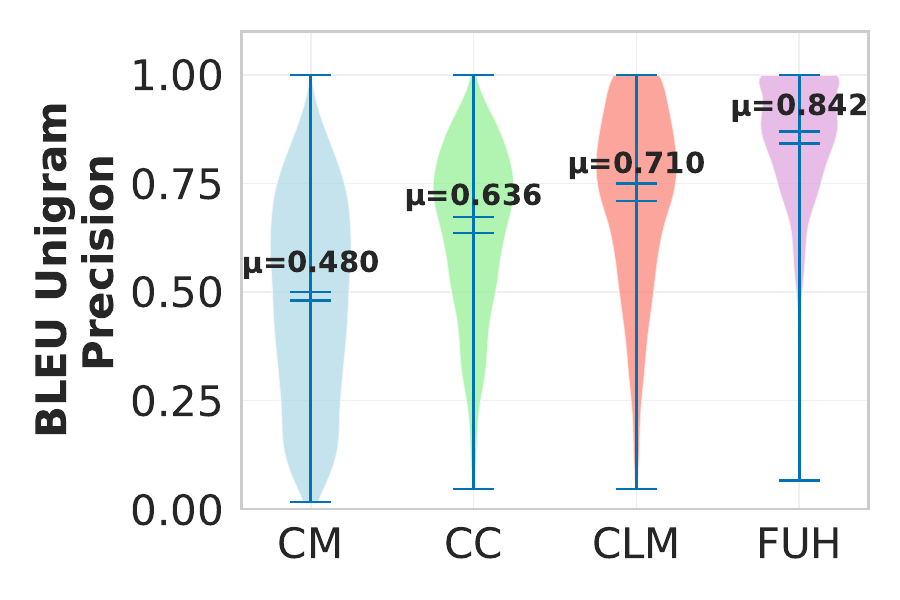}
        \caption{BLEU Unigram Precision}
        \label{fig: BleuScore}
    \end{subfigure}
    %\fi 
    \begin{subfigure}[t]{0.45\columnwidth}
        \centering
        \includegraphics[width=\linewidth]{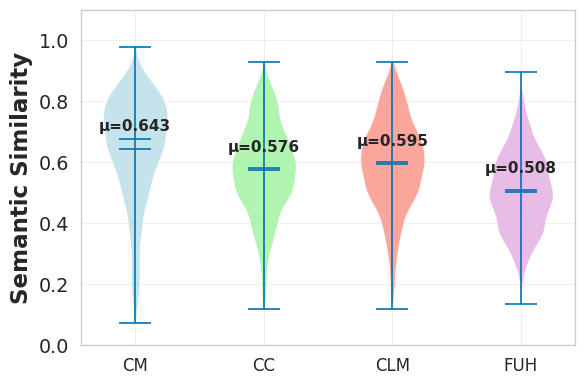}
         \caption{Cosine Similarity}
        \label{fig: BertScore}
    \end{subfigure}
    %\if 0 
    \begin{subfigure}[t]{0.45\columnwidth}
        \centering
        \includegraphics[width=\linewidth]{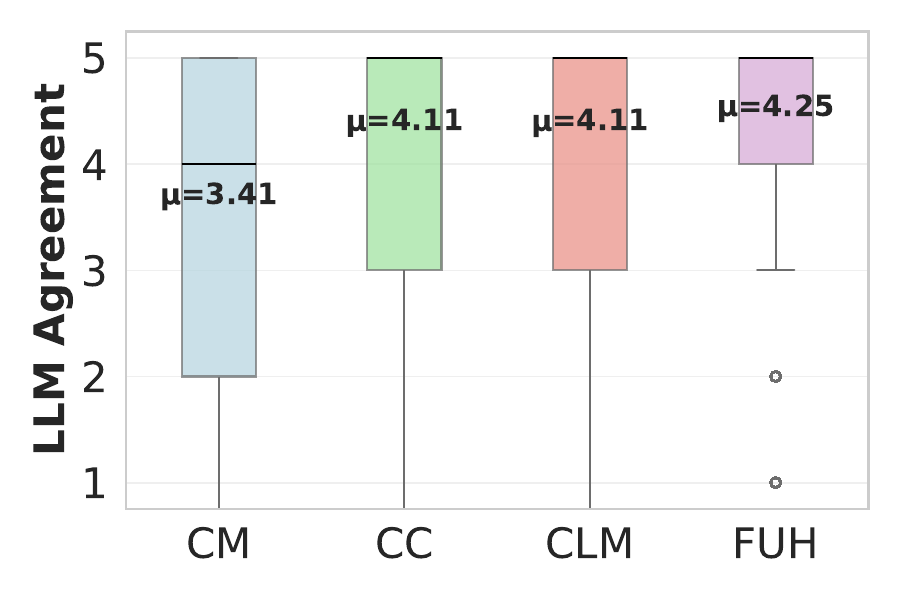}
         \caption{LLM Agreement Scores}
        \label{fig: LLMEval}
    \end{subfigure}
    %\fi 
    \caption{(a--b)~Syntactic, (c)~Semantic, and (d)~LLM evaluation based provenance measures. Majority of ChatGPT’s memories are directly or semantically grounded in user input.}
    \label{fig: provenance}
    % \vspace{-5mm}
\end{figure}

\noindent
\textbf{Direct Grounding in User Text:}
On average 84\% memory entries show direct string overlap (see rightmost distribution in \Cref{fig: EMRate}) when considering full user history. 
When we reduce the context to CLM, CC and CU the exact match rate reduces to 70\%, 63\%, and 47\% respectively. 
This observation suggests that a huge amount of memories in our dataset are directly grounded in participants' texts with the direct grounding already occurring locally (in the current conversation context) for more than half of them. 
The sharp increase in percentage when we add past conversation memories indicate that previously generated memories may influence the formation of subsequent ones. 
Evaluation using BLEU unigram precision (see \Cref{fig: BleuScore}) mirrors this trend as well.

\noindent
\textbf{Paraphrasing and Summarization:}
\Cref{fig: BertScore} reports the semantic similarity between memory entries and the specified user context.
\new{Semantic similarity remains consistently high across all context configurations ($\ge 0.51$). 
Such consistent high scores suggest that memory entries are semantically aligned with multiple representations of the surrounding conversation.} 

\noindent
\textbf{Inference:}
\Cref{fig: LLMEval} shows the distribution of the accuracy analyses using an LLM as a judge. 
Using the five-point Likert scale, $77\%$ of messages receive a score $\geq 4$ (``directly stated'' or ``paraphrased/summarized''), $14\%$ are rated as ``logically inferred'' (score 3), $5\%$ as ``weakly supported'', and $4\%$ as ``not supported'' considering the full user history (FUH) setting. 
Overall, for FUH, CLM, CC, and CM the mean accuracy rating is found to be $4.25$, $4.11$, $4.11$ and $3.4$ respectively. 
\new{To assess the reliability of LLM evaluations, two authors annotated the combinations of 10 randomly sampled memories from each scores (50 in total). 
Annotators achieved an almost perfect agreement (99\%) with average agreement of 93.7\% with the LLM across combinations.}

LLM-based evaluations further indicate that most of the memories are grounded in the user context. 
During the manual annotation process authors found two potential causes for lower ratings by LLMs. 
Firstly, some memories are generated when the participants might have uploaded some files containing texts -- which was not shown to the LLM while evaluating. 
Secondly, on some occasions ChatGPT might have slightly extrapolated the user intent within the message while generating the memories. 
For example, in a case where a participant asks for ``cheap meal suggestions [...] for more energy throughout their day'', the corresponding memory is ``...wants to improve his energy levels by adjusting his diet''. 
The LLM categorizes this as logically inferred.
In a context where the participant specifies ``Give me bands like Nirvana'', the created memory is ``User likes Nirvana.'', which is not explicitly stated and relies on an assumption that Nirvana implies liking them. The model annotates this as weakly supported. 
\Cref{tab:examples-provenance} shows more examples for LLM evaluations in CM setting. 

\begin{table*}[t!]
\caption{Anecdotal samples corresponding to LLM agreement scores on the provenance of memories for \textit{CM} setting.}
\label{tab:examples-provenance}
\small
\centering
\begin{tabular}{|l|l|l|}
\hline
\textbf{LLM Agreement} & \textbf{User Message}                                 & \textbf{Memory} \\ \hline
%1 = Unsupported/Contradicted                 & \begin{tabular}[c]{@{}l@{}}"Well, this is the link towards the product, so you go ahead \\ and analyze that and take whatever information that \\ I haven't told you about and see if you can come up \\ with how many servings are per bottle. \\ XX to Reduce Stress and Promote Relaxation.."\end{tabular} & \begin{tabular}[c]{@{}l@{}}"User's doctor said that the pain \\ in the ankle and foot is probably \\ related to their sleeping position and \\ suggested doing some ankle stretches."\end{tabular}                  \\ \hline
1 = Unsupported/Contradicted                  & \begin{tabular}[c]{@{}l@{}} "Please write a cover letter for this job focusing \\ on my experience working in XX and \\ my voluntary experience working with XX.." \\ *\textit{<File was uploaded>} \end{tabular}                                                  & "The user's name is XXX XXXX."                  \\ \hline
2 = Weakly Supported                 & \begin{tabular}[c]{@{}l@{}} "Change XXX occupation to something \\ more interesting. Not travel writer \\ and photographer and nothing VR related"     \end{tabular}                                                 & \begin{tabular}[c]{@{}l@{}} "XXX's occupation has been \\ changed to a creative director \\ for a sustainable fashion brand."    \end{tabular}                  \\ \hline
3 = Logically Inferred                  & \begin{tabular}[c]{@{}l@{}} "What does schopenhauer mean  by: “That which \\ knows all things and is known by none is the subject."      \end{tabular}                                                       & \begin{tabular}[c]{@{}l@{}} "The user is exploring Schopenhauer's \\ concept of the subject and its \\implications for knowledge."     \end{tabular}                  \\ \hline
4 = Paraphrased/Summarized                & \begin{tabular}[c]{@{}l@{}} "I am preparing for a job interview for XX \\ I need to prepare questions to ask my \\ potential employer at the end of the interview. \\ Can you help me with preparing these questions?     \end{tabular}                                                       & \begin{tabular}[c]{@{}l@{}} "X is preparing for a job \\ interview for XX position."     \end{tabular}                  \\ \hline
5 = Directly stated                 & \begin{tabular}[c]{@{}l@{}} "I already reverted the senate's ability to dissolve \\ the house in this model"     \end{tabular}                                                       & \begin{tabular}[c]{@{}l@{}}  "User has reverted the Senate's ability \\ to dissolve the House in their model."    \end{tabular}                  \\ \hline
\end{tabular}
\end{table*}

Our provenance analyses show that %while 
a significant majority of ChatGPT’s memories are directly or semantically grounded in user inputs. %, a nontrivial fraction are inferred or weakly supported.
While in some cases memory entries are summarized or logically inferred, they faithfully represent most of the conversations between participants and the conversational system.

\section{Reverse Engineering Memories}~\label{Sec: Tool}

\begin{table*}[t!]
\caption{\textit{Top}: Syntactic and semantic evaluation of memories; \textit{Bottom}: Anecdotal samples of predicted memories for the given user query using \texttt{Qwen2.5-32B} fine-tuned and ICL model.}
\footnotesize
%\smaller
\begin{tabular}{|l|llllll|llllll|llllll|}
\hline
\multirow{4}{*}{\textbf{Models}}                         & \multicolumn{6}{c|}{\textbf{Ground Truth}}                                                                                               & \multicolumn{6}{c|}{\textbf{User Query}}                                                                                                 & \multicolumn{6}{c|}{\textbf{Context + User Query}}          
\\ \cline{2-19} 
                                                         & \multicolumn{2}{l|}{\makecell{BLEU\\Recall}
                                                         }                  & \multicolumn{2}{l|}{\makecell{ROUGE\\Recall}}                         & \multicolumn{2}{l|}{\makecell{Semantic\\ Similarity}} & \multicolumn{2}{l|}{\makecell{BLEU\\Precision}}                          & \multicolumn{2}{l|}{\makecell{ROUGE\\Precision}}                         & \multicolumn{2}{l|}{\makecell{Semantic\\ Similarity}} & \multicolumn{2}{l|}{\makecell{BLEU\\Precision}}                          & \multicolumn{2}{l|}{\makecell{ROUGE\\Precision}}                         & \multicolumn{2}{l|}{\makecell{Semantic\\ Similarity}} 
                                                         \\ \cline{2-19} %&
%\multicolumn{4}{c|}{Recall}& \multicolumn{2}{c|}{F1} & \multicolumn{4}{c|}{Precision}& \multicolumn{2}{c|}{F1} & \multicolumn{4}{c|}{Precision}& \multicolumn{2}{c|}{F1} 
                                                         %\\ \cline{2-19} 
                                                         & \multicolumn{1}{l|}{ICL} & \multicolumn{1}{l|}{FT} & \multicolumn{1}{l|}{ICL} & \multicolumn{1}{l|}{FT} & \multicolumn{1}{l|}{ICL}  & FT & \multicolumn{1}{l|}{ICL} & \multicolumn{1}{l|}{FT} & \multicolumn{1}{l|}{ICL} & \multicolumn{1}{l|}{FT} & \multicolumn{1}{l|}{ICL}  & FT & \multicolumn{1}{l|}{ICL} & \multicolumn{1}{l|}{FT} &\multicolumn{1}{l|}{ICL} & \multicolumn{1}{l|}{FT} & \multicolumn{1}{l|}{ICL} & \multicolumn{1}{l|}{FT}\\ \hline
\begin{tabular}[c]{@{}l@{}}Qwen2.5\\ 32B-it\end{tabular} & \multicolumn{1}{l|}{0.41} & \multicolumn{1}{l|}{0.43} & \multicolumn{1}{l|}{0.35} & \multicolumn{1}{l|}{0.39} & \multicolumn{1}{l|}{\new{0.68}} & \multicolumn{1}{l|}{\new{0.70}} & \multicolumn{1}{l|}{0.27} & \multicolumn{1}{l|}{0.36} & \multicolumn{1}{l|}{0.22} & \multicolumn{1}{l|}{0.31} & \multicolumn{1}{l|}{\new{0.66}} & \multicolumn{1}{l|}{\new{0.63}} & \multicolumn{1}{l|}{0.53} & \multicolumn{1}{l|}{0.56} & \multicolumn{1}{l|}{0.41} & \multicolumn{1}{l|}{0.45} & \multicolumn{1}{l|}{\new{0.69}} & \multicolumn{1}{l|}{\new{0.59}} \\ \hline
\begin{tabular}[c]{@{}l@{}}Gemma3\\ 27B-it\end{tabular}  & \multicolumn{1}{l|}{0.39} & \multicolumn{1}{l|}{0.31} & \multicolumn{1}{l|}{0.32} & \multicolumn{1}{l|}{0.28} & \multicolumn{1}{l|}{\new{0.62}} & \multicolumn{1}{l|}{\new{0.59}} & \multicolumn{1}{l|}{0.23} & \multicolumn{1}{l|}{0.34} & \multicolumn{1}{l|}{0.18} & \multicolumn{1}{l|}{0.29} & \multicolumn{1}{l|}{\new{0.62}} & \multicolumn{1}{l|}{\new{0.53}} & \multicolumn{1}{l|}{0.44} & \multicolumn{1}{l|}{0.50} & \multicolumn{1}{l|}{0.32} & \multicolumn{1}{l|}{0.42} & \multicolumn{1}{l|}{\new{0.66}} & \multicolumn{1}{l|}{\new{0.39}} \\ \hline
\begin{tabular}[c]{@{}l@{}}GPT-OSS\\ 20B\end{tabular}    & \multicolumn{1}{l|}{0.19} & \multicolumn{1}{l|}{0.35} & \multicolumn{1}{l|}{0.16} & \multicolumn{1}{l|}{0.32} & \multicolumn{1}{l|}{\new{0.49}} & \multicolumn{1}{l|}{\new{0.65}} & \multicolumn{1}{l|}{0.20} & \multicolumn{1}{l|}{0.35} & \multicolumn{1}{l|}{0.18} & \multicolumn{1}{l|}{0.31} & \multicolumn{1}{l|}{\new{0.55}} & \multicolumn{1}{l|}{\new{0.58}} & \multicolumn{1}{l|}{0.37} & \multicolumn{1}{l|}{0.52} & \multicolumn{1}{l|}{0.32} & \multicolumn{1}{l|}{0.44} & \multicolumn{1}{l|}{\new{0.50}} & \multicolumn{1}{l|}{\new{0.52}} \\ \hline
\end{tabular}

%---- Subfigure 1 ----
    \begin{minipage}[b]{0.24\textwidth}
        \centering
        \includegraphics[width=\textwidth]{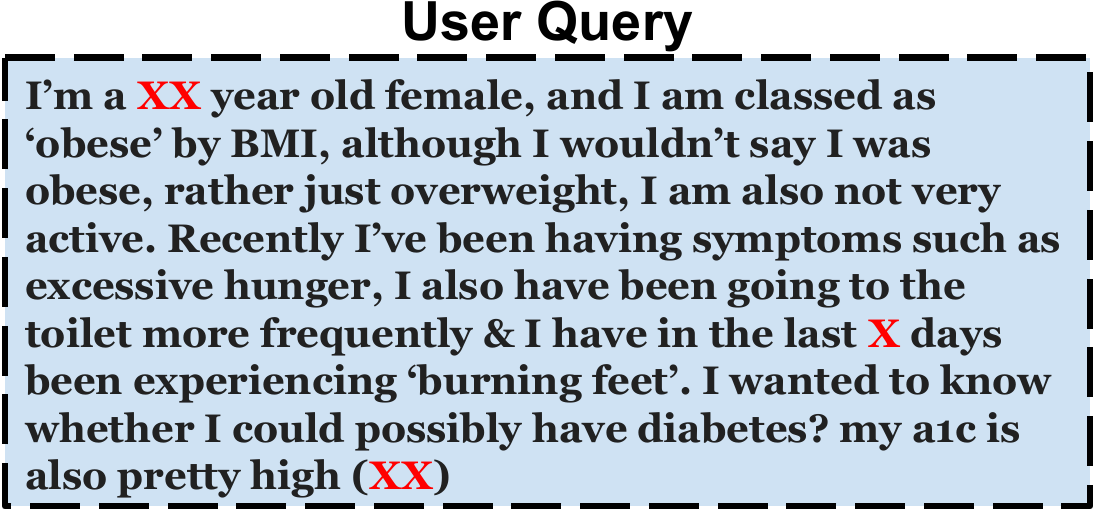}
        % \caption*{(a) Caption 1}
    \end{minipage}
    \hfill
    %---- Subfigure 2 ----
    \begin{minipage}[b]{0.24\textwidth}
        \centering
        \includegraphics[width=\textwidth]{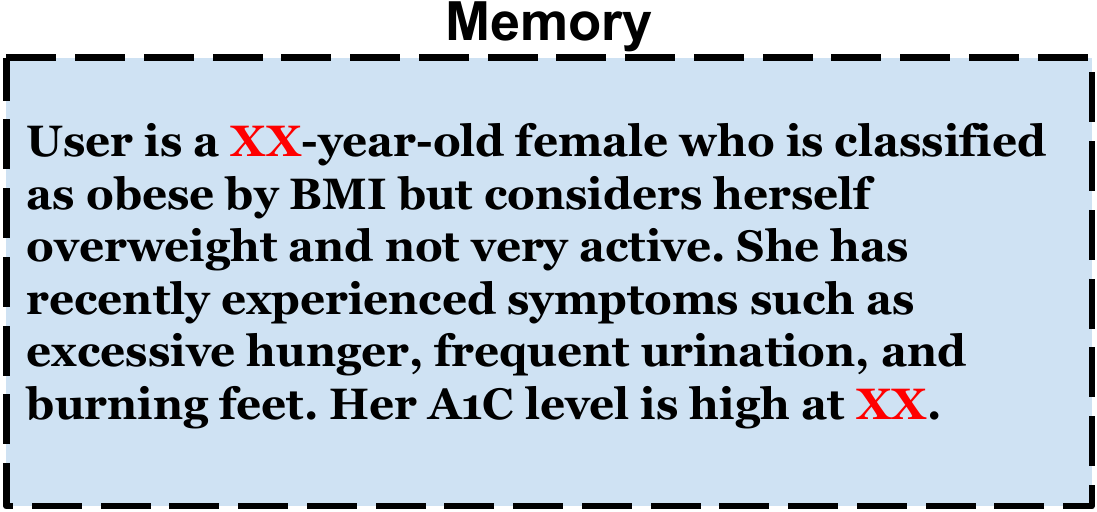}
        % \caption*{(b) Caption 2}
    \end{minipage}
    \hfill
    %---- Subfigure 3 ----
    \begin{minipage}[b]{0.24\textwidth}
        \centering
        \includegraphics[width=\textwidth]{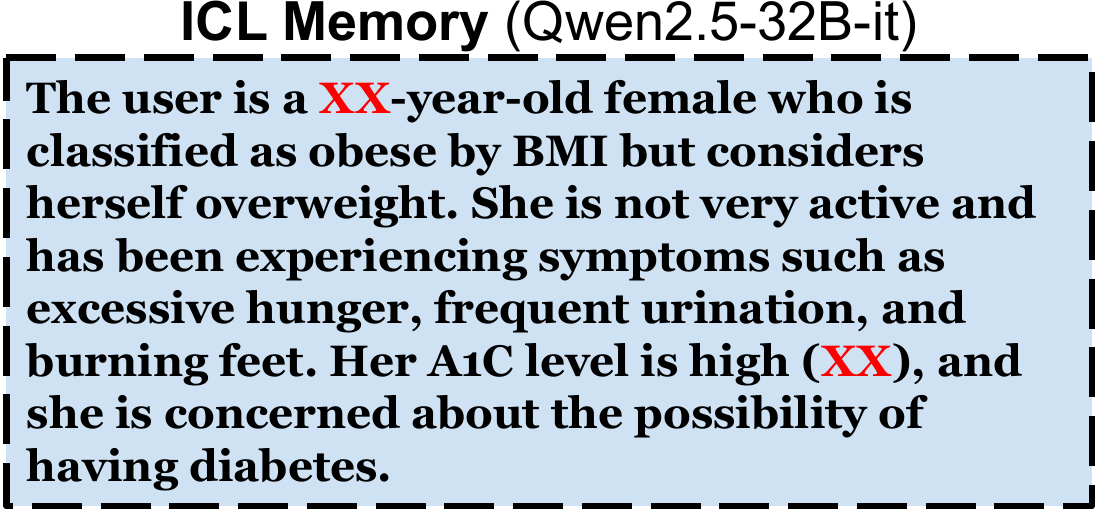}
        % \caption*{(c) Caption 3}
    \end{minipage}
    \hfill
    %---- Subfigure 4 ----
    \begin{minipage}[b]{0.24\textwidth}
        \centering
        \includegraphics[width=\textwidth]{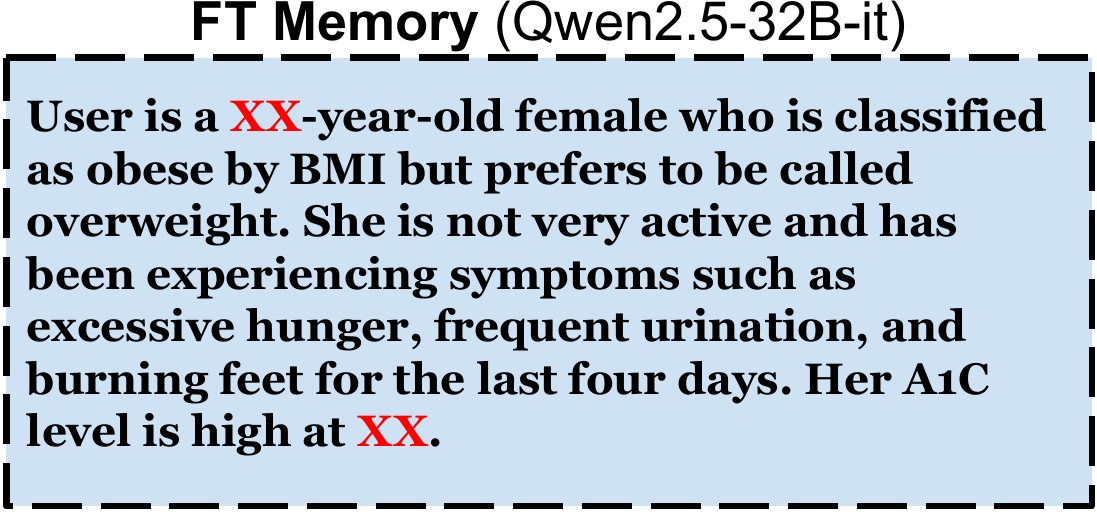}
        % \caption*{(d) Caption 4}
    \end{minipage}

\label{tab:syn-sem-eval-memory-eng}
%\vspace{-7mm}
\end{table*}

\begin{table*}[t!]
\caption{\textit{Top}: Syntactic and semantic evaluation of rephrased queries; \textit{Bottom}: Anecdotal samples of predicted rephrased queries for the given user query using \texttt{Qwen2.5-32B} fine-tuned and ICL model.}
%\scriptsize
\footnotesize
%\smaller
\begin{tabular}{|l|llllll|llllll|llllll|}
\hline
\multirow{4}{*}{\textbf{Models}}                         & \multicolumn{6}{c|}{\textbf{Ground Truth}}                                                                                               & \multicolumn{6}{c|}{\textbf{User Query}}                                                                                                 & \multicolumn{6}{c|}{\textbf{Context + User Query}}          
\\ \cline{2-19} 
                                                         & \multicolumn{2}{l|}{\makecell{BLEU\\Recall}}                          & \multicolumn{2}{l|}{\makecell{ROUGE\\Recall}}                         & \multicolumn{2}{l|}{\makecell{Semantic\\Similarity}} & \multicolumn{2}{l|}{\makecell{BLEU\\Precision}}                          & \multicolumn{2}{l|}{\makecell{ROUGE\\Precision}}                         & \multicolumn{2}{l|}{\makecell{Semantic\\Similarity}} & \multicolumn{2}{l|}{\makecell{BLEU\\Precision}}                          & \multicolumn{2}{l|}{\makecell{ROUGE\\Precision}}                         & \multicolumn{2}{l|}{\makecell{Semantic\\Similarity}} 
                                                         \\ \cline{2-19} %&
%\multicolumn{4}{c|}{Recall}& \multicolumn{2}{c|}{F1} & \multicolumn{4}{c|}{Precision}& \multicolumn{2}{c|}{F1} & \multicolumn{4}{c|}{Precision}& \multicolumn{2}{c|}{F1} 
                                                         %\\ \cline{2-19} 
                                                         & \multicolumn{1}{l|}{ICL} & \multicolumn{1}{l|}{FT} & \multicolumn{1}{l|}{ICL} & \multicolumn{1}{l|}{FT} & \multicolumn{1}{l|}{ICL}  & FT & \multicolumn{1}{l|}{ICL} & \multicolumn{1}{l|}{FT} & \multicolumn{1}{l|}{ICL} & \multicolumn{1}{l|}{FT} & \multicolumn{1}{l|}{ICL}  & FT & \multicolumn{1}{l|}{ICL} & \multicolumn{1}{l|}{FT} &\multicolumn{1}{l|}{ICL} & \multicolumn{1}{l|}{FT} & \multicolumn{1}{l|}{ICL} & \multicolumn{1}{l|}{FT}\\ \hline
\begin{tabular}[c]{@{}l@{}}Qwen2.5\\ 32B-it\end{tabular} & \multicolumn{1}{l|}{0.16} & \multicolumn{1}{l|}{0.09} & \multicolumn{1}{l|}{0.13} & \multicolumn{1}{l|}{0.08} & \multicolumn{1}{l|}{\new{0.55}} & \multicolumn{1}{l|}{\new{0.42}} & \multicolumn{1}{l|}{0.22} & \multicolumn{1}{l|}{0.14} & \multicolumn{1}{l|}{0.18} & \multicolumn{1}{l|}{0.12} & \multicolumn{1}{l|}{\new{0.59}} & \multicolumn{1}{l|}{\new{0.47}} & \multicolumn{1}{l|}{0.40} & \multicolumn{1}{l|}{0.24} & \multicolumn{1}{l|}{0.32} & \multicolumn{1}{l|}{0.20} & \multicolumn{1}{l|}{\new{0.53}} & \multicolumn{1}{l|}{\new{0.46}} \\ \hline
\begin{tabular}[c]{@{}l@{}}Gemma3\\ 27B-it\end{tabular}  & \multicolumn{1}{l|}{0.21} & \multicolumn{1}{l|}{0.11} & \multicolumn{1}{l|}{0.17} & \multicolumn{1}{l|}{0.10} & \multicolumn{1}{l|}{\new{0.46}} & \multicolumn{1}{l|}{\new{0.36}} & \multicolumn{1}{l|}{0.23} & \multicolumn{1}{l|}{0.13} & \multicolumn{1}{l|}{0.19} & \multicolumn{1}{l|}{0.11} & \multicolumn{1}{l|}{\new{0.52}} & \multicolumn{1}{l|}{\new{0.38}} & \multicolumn{1}{l|}{0.38} & \multicolumn{1}{l|}{0.26} & \multicolumn{1}{l|}{0.30} & \multicolumn{1}{l|}{0.22} & \multicolumn{1}{l|}{\new{0.47}} & \multicolumn{1}{l|}{\new{0.37}} \\ \hline
\begin{tabular}[c]{@{}l@{}}GPT-OSS\\ 20B\end{tabular}    & \multicolumn{1}{l|}{0.11} & \multicolumn{1}{l|}{0.09} & \multicolumn{1}{l|}{0.10} & \multicolumn{1}{l|}{0.08} & \multicolumn{1}{l|}{\new{0.33}} & \multicolumn{1}{l|}{\new{0.46}} & \multicolumn{1}{l|}{0.09} & \multicolumn{1}{l|}{0.09} & \multicolumn{1}{l|}{0.08} & \multicolumn{1}{l|}{0.07} & \multicolumn{1}{l|}{\new{0.43}} & \multicolumn{1}{l|}{\new{0.46}} & \multicolumn{1}{l|}{0.24} & \multicolumn{1}{l|}{0.16} & \multicolumn{1}{l|}{0.19} & \multicolumn{1}{l|}{0.13} & \multicolumn{1}{l|}{\new{0.40}} & \multicolumn{1}{l|}{\new{0.44}} \\ \hline
\end{tabular}

%---- Subfigure 1 ----
\begin{minipage}[b]{0.24\textwidth}
    \centering
    \includegraphics[width=\textwidth]{Figures/examples/query.pdf}
    % \caption*{(a) Caption 1}
\end{minipage}
\hfill
%---- Subfigure 2 ----
\begin{minipage}[b]{0.24\textwidth}
    \centering
    \includegraphics[width=\textwidth]{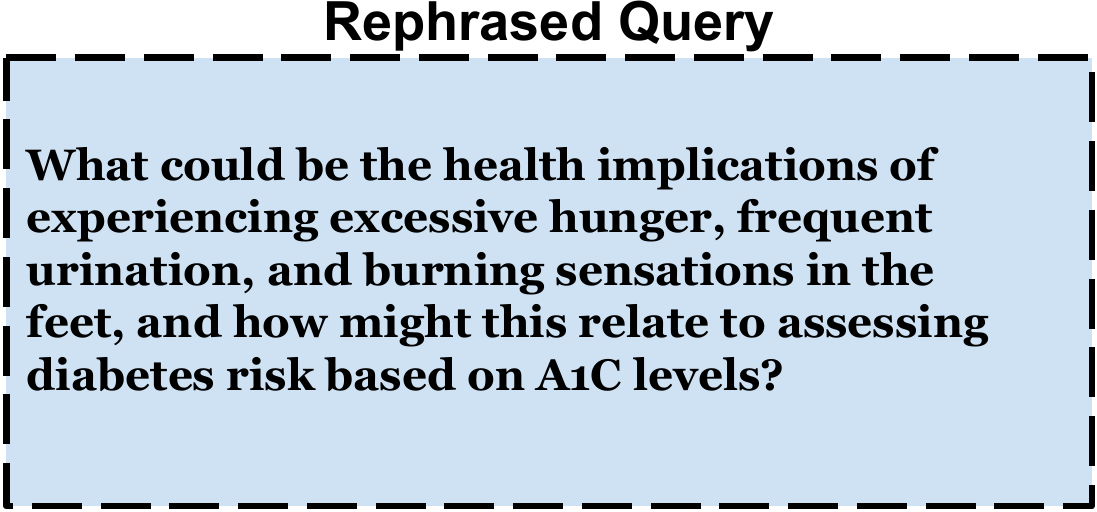}
    % \caption*{(b) Caption 2}
\end{minipage}
\hfill
%---- Subfigure 3 ----
\begin{minipage}[b]{0.24\textwidth}
    \centering
    \includegraphics[width=\textwidth]{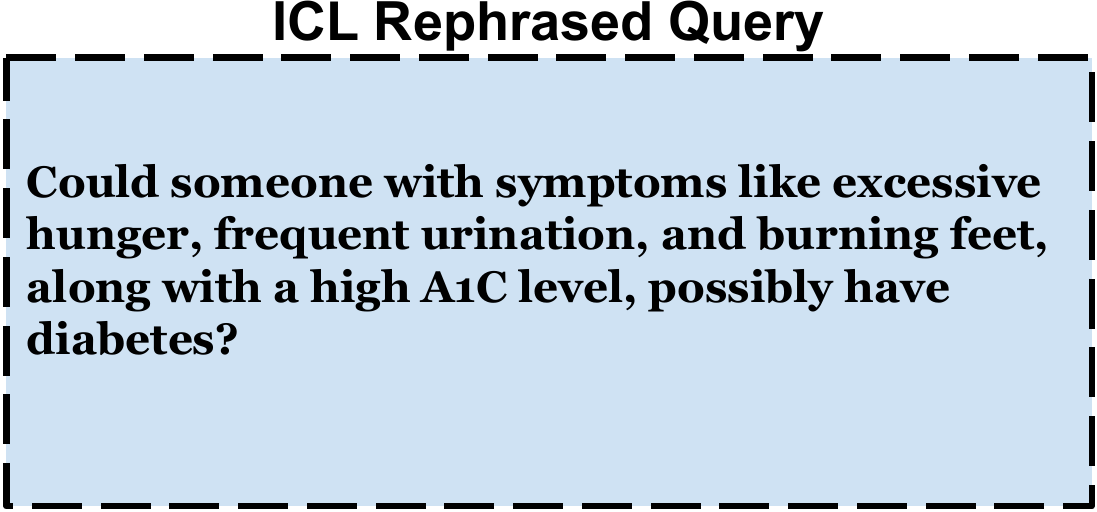}
    % \caption*{(c) Caption 3}
\end{minipage}
\hfill
%---- Subfigure 4 ----
\begin{minipage}[b]{0.24\textwidth}
    \centering
    \includegraphics[width=\textwidth]{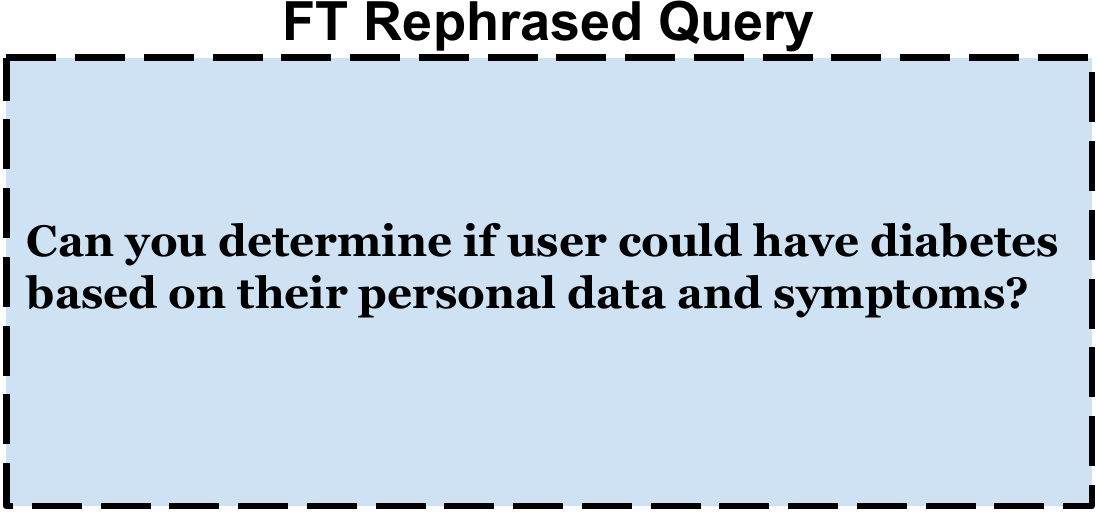}
    % \caption*{(d) Caption 4}
\end{minipage}

\label{tab:syn-sem-eval-rephrase-eng}
%\vspace{-5mm}
\end{table*}

\new{In \Cref{subsec:agencyuser}, we observe that ChatGPT can implicitly store sensitive information from user queries in its memory, indicating a gap between policy and practice. 
Such gap underscores the need for a system that (a)~alerts users about the extent of sensitive information that can be inferred from their query,   
(b)~recommends them a reformulated query that  
prevents exposition of the sensitive information, while retaining the original intent of their query. 
To this end, in this section, we introduce a framework with the goal of (i)~imitating the memory extractor, (ii)~estimating potential risks of extracted memories related to sensitive data, and (iii)~recommending rephrased queries 
to minimize the attribution to sensitive data in memories, thus being more compliant with the policies.}

\subsection{Experimental Framework}
\label{subsec:expsetup}
Next, we describe our framework, including the dataset, models, and evaluation metrics. 

\vspace{1mm}
\noindent
\textbf{Dataset:} %Following Section~\ref{Sec: Dataset}, 
We use $22,971$ participant queries ($2,050$ with memory entries) and prompt GPT-4o to identify personal data (if any) and produce corresponding rephrased queries.
The rephrased queries retain user intent while avoiding personal attribution, e.g., a user query like
\textit{``I really need to quit smoking cannabis"} is rephrased as \textit{``What are some strategies for quitting cannabis?'' } 
We compiled a dataset with a total of $14,834$ queries, where we retained the original $2,050$ memory entries (with or without rephrased query), and all the remaining $12,784$ queries have rephrased queries. The drop from $22,971$ to $14,834$ is because we removed all queries that lacked rephrased queries indicating the queries did \textit{not} contain any sensitive data.
Each data point includes the user query, its conversational context, the original memory (if present), and the rephrased query. 
We split the data assigning 60\% of each user’s memory queries to training and the remaining to testing. 

\vspace{1mm}
\noindent
\textbf{Models:} We use \texttt{Qwen2.5-32B-it} \citep{qwen2025qwen25technicalreport}, \texttt{Gemma3-27B-it} \citep{team2025gemma}, and \texttt{GPT-OSS-20B} \citep{agarwal2025gpt} for fine-tuning (FT) and in-context learning (ICL). 
However, since training models incrementally on each incoming user's data is impractical,
we train on data from $5$ randomly sampled users and evaluate on the full test set ($\sim13.6k$ queries) over all available users. 
FT uses all the training data from the $5$ users ($\sim 87$), while ICL uses $10$ in-context samples (2 samples/ user). 
\new{We provide the accompanying prompts and instructions in the code release.}

\vspace{1mm}
\noindent
\textbf{Metrics:} We evaluate syntactic similarity using BLEU-4~\citep{papineni2002bleu} and ROUGE-L~\citep{lin2004rouge}, and semantic similarity \new{using cosine similarity on text embeddings with \texttt{openai/text-embedding-3-large}}.
%BERTScore~\citep{zhang2019bertscore} with XLM-RoBERTa-Large~\citep{conneau2019unsupervised}. 
Predictions are compared against (i) ground truth, (ii) user queries, and (iii) context + user queries, reflecting how ChatGPT memories may span across inputs (as seen in Section \ref{Sec : Accuracy}). For syntactic metrics, we report recall with respect to the ground truth, as overlaps between predicted memories and user inputs or context can reduce precision. \new{Precision is reported for syntactic metrics against user queries and context, and cosine similarity for the semantic metric.
}

%\vspace{-8mm}
\subsection{Imitating Memory Extractor}
\label{subsec:memextract}

Given a user query and its conversational context (past queries), our tool predicts the memory that could have been extracted. Predicted memories are evaluated against (i) ground truth, (ii) user queries, and (iii) context + user queries to assess similarity. \new{These $20-32$B models roughly take $130-390$ ms per query, indicating good scalability.}
Table~\ref{tab:syn-sem-eval-memory-eng} reports the performance of FT and ICL models across three model families for English-dominant users ($35$ users, $\sim6.9$k queries). Predictions show \new{reasonable} semantic similarity, with scores following the trend as in Figure \ref{fig: BertScore}: ground truth $>$ user query $>$ context + user query. Syntactic scores are higher when compared with context, consistent with Figures~\ref{fig: EMRate} and~\ref{fig: BleuScore}.
For non–English-dominant users ($30$ users, $\sim6.7$k queries; Table~\ref{tab:syn-sem-eval-memories-noneng} in Appendix~\ref{app:results_pg}), a similar pattern emerges but syntactic scores are lower, as predictions are generated in English while original memories are in other languages (e.g., original: \textit{``L'utente ha l'artrite reumatoide."}, prediction: \textit{``User has rheumatoid arthritis"}). Semantic scores remain high, \new{due to the inherent multilingual property of the open-source models.}
Overall, FT and ICL perform comparably, with FT occasionally outperforming ICL, highlighting the complementary roles of embedded and external knowledge. 
Bottom panel of Table~\ref{tab:syn-sem-eval-memory-eng} illustrates an anecdotal example affirming the memory extractor's imitation capability. 
\new{Two authors evaluated FT and ICL memories on a 5-point Likert scale (1: very dissimilar, 5: very similar), finding substantial agreement (73.5\%) and high similarity (rated > 4) to ground truth for 77.77\% (FT) and 83.83\% (ICL) of cases.}

\subsection{Estimating Potential Risks}
\label{subsec:estimaterisk}

\begin{figure}[t]
    \centering
    \begin{subfigure}[t]{0.44\columnwidth}
        \centering
        \includegraphics[width=\linewidth]{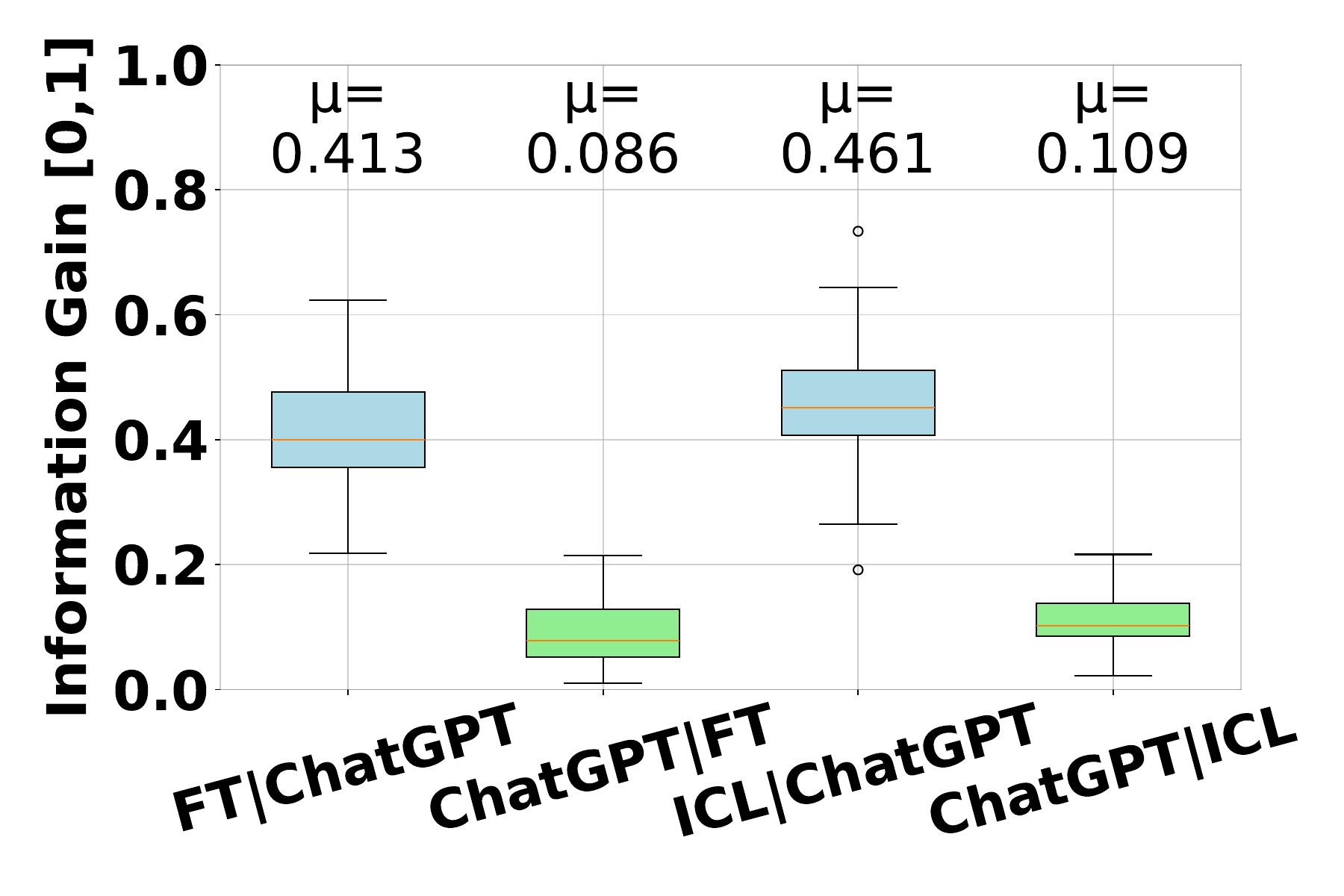}
        %\caption{Preventing Attribution}
        \caption{Information gain}
        \label{fig:risks-mem}
        %\label{fig:privacy_attrib_ft}
    \end{subfigure}
    \hfill
    \begin{subfigure}[t]{0.49\columnwidth}
        \centering
        \includegraphics[width=\linewidth]{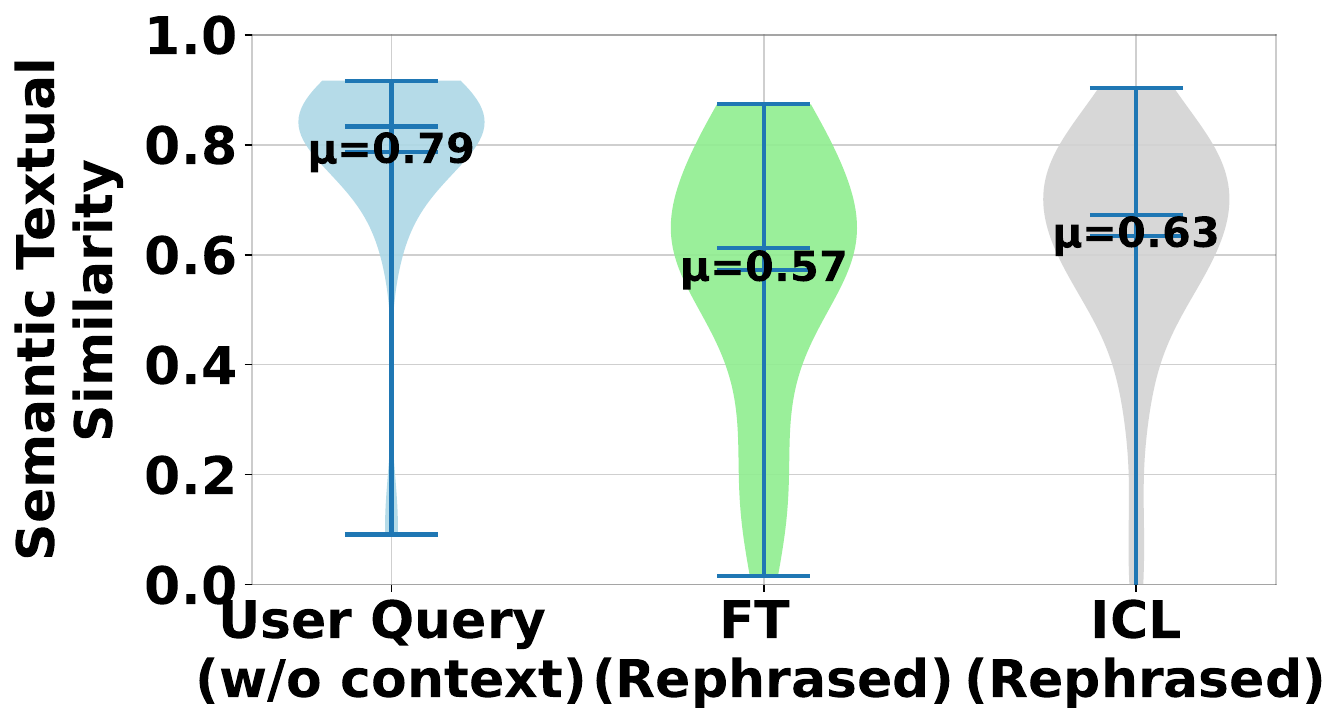}
        \caption{Utility of rephrased queries}
        \label{fig:utility_rephrased}
    \end{subfigure}
    \vspace{-1mm}
    \caption{
    (a)~Information gain of all predicted memories over ChatGPT memories shows high gain, hinting at higher potential risks.
    (b)~Responses from %(1)~user query without context, and (2)~
    rephrased queries from ICL and FT are semantically similar to the original response. %leading to high BERTScore.
    }
    \label{fig:privacy_attribution_comparison}
    %\vspace{-5mm}
\end{figure}

ChatGPT currently extracts memories from specific user conversations, and the criteria for saving memories (which conversations or how many) may vary over time. %If memories were saved for all user queries, the scenario would change. 
In this context, an interesting question to understand the landscape of privacy threat would be : \textit{if ChatGPT were to extract memory for all user queries}, %We are interested in quantifying 
\textit{how much additional information such all-possible extracted memories could hold} compared to those collected by ChatGPT today. To quantify this, we use information gain, which measures how much new semantic content text $Y$ \new{(extracted memories)} adds beyond text $X$ \new{(original memories)}, denoted as $Y|X$. Inspired from \citep{le2025filtering} which introduces semantic novelty, we define embedding-based information gain (IG) as the complement of the average maximal cosine similarity:\\
%\begin{equation}
%\nonumber
    $IG(Y|X) = 1- \frac{1}{|Y|} \sum _{y \in Y} max _{x \in X}[max(0, cos(e(y), e(x)))]$
%\end{equation}
where $e(.)$ is the sentence embedding using All-MiniLM-L6-v2 model \citep{reimers2021allminilm}. Figure~\ref{fig:risks-mem} shows that the IG of all extracted memories from our FT and ICL models over ChatGPT is significantly higher ($\sim 0.46$) than that of ChatGPT over them ($\sim 0.1$). \new{We used GPT-4o to evaluate sensitive content in extracted memories and observed that, while only 28\% of original memories contained sensitive information, this increased to 35\% (ICL) and 31.64\% (FT Qwen) in the extracted memories, %highlighting potential risks such as sensitive data leakage when storing all memories. 
indicating storage of higher amount of sensitive information about users. 
We also analyzed Theory of Mind (ToM) content and found that its presence rose from 52\% in original memories to 59.1\% (ICL) and 55.4\% (FT) in extracted memories.}

\subsection{Attribution Shield}
\label{sec:attributionshield}
Having observed the potential risks of memories elicited from user queries, we design a risk mitigation strategy that recommends rephrased queries, while preserving user intent. By risk mitigation, we refer to the tool's ability to \textit{shield/prevent user attribution} in rephrased queries. We evaluate its effectiveness in generating rephrased queries and their utility in producing relevant responses.  

\noindent
\textbf{Preventing attribution in rephrased queries: }
%\label{subsub:privacy-query}
Given a user query and its context, the tool predicts a rephrased query that is generic and identity-preserving. Table~\ref{tab:syn-sem-eval-rephrase-eng} reports the performance of FT and ICL models across $3$ model families for English-dominant users. Syntactic scores are modest, as rephrased queries are intentionally generic, but semantic similarity remains reasonably good across the board. Similar patterns are observed for non-English-dominant users (Table~\ref{tab:syn-sem-eval-memories-noneng} (lower panel) in Appendix~\ref{app:results_pg}), demonstrating the model’s ability to preserve query intent while preventing attribution (see Table~\ref{tab:syn-sem-eval-rephrase-eng} bottom panel for anecdotal example).

To quantify prevention of user attribution, we select $\sim36$ user queries that yield memories on first occurrence. This strategy helps us in avoiding the replay of the entire conversation. We also have predicted rephrased queries for those user queries from both our fine-tuned and ICL based models.  We present each \textit{<user query, rephrased query>} pair to GPT-4o, asking: `\textit{Which of them is more privacy preserving? We define privacy preserving in terms of less attribution to personal actions.}'. We observe that that $94.4\%$ of FT rephrased queries and 100\% of ICL rephrased queries are attributing less to the users than the originals, confirming their effectiveness. \new{After two authors annotating the rephrased queries, they found that 100\% (FT) and 95.8\% (ICL) of the rephrased queries were more privacy preserving with $83\%$ agreement between them.
}

\noindent
\textbf{Utility of rephrased user queries: }
%\label{subsub:utility-rephrased}
Using the same $36$ queries, we compare 3 responses from ChatGPT: (i) original response including context, (ii) response to the original query without context, and (iii) response to the rephrased queries. We measure semantic textual similarity between each generated response and the original response. We provide the prompts to get the response from the OpenAI API in the \new{code release}. Figure~\ref{fig:utility_rephrased} shows comparable similarity, indicating that neither the omission of context nor the anonymity of rephrased queries affects the responses significantly. Anecdotal examples in Figure~\ref{tab:anec-responses} illustrate the consistency of responses. \new{After annotation by two authors, we found that utility was preserved in 91\% (ICL) and 87\% (FT) of rephrased queries with $94\%$ agreement.}
\section{Concluding Discussion}\label{Sec: Discussion}

In this paper, we investigate ChatGPT's memory feature, conceptualizing it as an `algorithmic self-portrait'. 
Our investigations reveal that these portraits are constructed with higher algorithmic autonomy, and capture a user's deep psychological framework : their `characteristic adaptations', raising profound security and privacy concerns. 
In response, we introduce \emph{Attribution Shield}, that reverse--engineers the memory generation process to alert users about potential sensitive memories and recommend reformulated queries to protect personal attribution while preserving utility.

\new{
At the same time, our findings should be considered in light of some limitations as well.
Firstly, our dataset, while ecologically valid, is sourced from a small set of Prolific users primarily in the US and Europe. 
The behaviors, and privacy attitudes of this group may not generalize to the global user base of conversational AI systems. 
However, given the rising concern of companionship usage of these conversational systems~\cite{karnam2026bowling}, we believe these findings are crucial for understanding and improving safety of human-AI interactions. }

\new{
Secondly, we study OpenAI's implementation of memory in ChatGPT. 
While the current implementation may change with time, and across platforms, the emerging security and privacy concerns are broadly applicable.  
Furthermore, the methodologies adopted across the different parts of the paper are generalizable to any other conversational systems and memory implementation. 
Finally, our analysis relies on state-of-the-art, but imperfect methodological proxies, including LLMs-as-judges. 
To this end, although we validated their results with human oversight, these tools may not fully capture the nuance of human language and intent.
}

\noindent \textbf{Ethical considerations:} 
This study was conducted with careful attention to ethical considerations and is consistent with the Ethical Review Board (ERB) of our university. % to conduct this research.
All data were obtained through GDPR-based data donations, with participants providing explicit informed consent to share their ChatGPT data for research purposes. 
The donated datasets were stored on secure servers and were neither shared with any third party nor will be released publicly due to their inherent sensitive nature.
We will delete the data within 3 years of completion of this study. 
For some of the analyses, we used GPT-4o as a judge to scale up annotations. 
The choice is motivated by two important considerations : (a)~GPT-4o is a model from OpenAI and is the model with whom participants had originally had the conversations, (b)~we leverage the EDU workspace of OpenAI which provides strict data protection and restrictions for usage of this data for training OpenAI models. 
For \textit{Attribution Shield}, all models are deployed on our institute's secured servers that abide by rigorous data protection and access control principles. 
 
Our findings highlight a fundamental shift in personalization, evolving the core challenge from traditional data privacy to the integrity of the algorithmic self-portrait. 
Such shift demands a new paradigm for both design and regulation: practitioners must build in--context tools that grant users real--time agency over how they are portrayed, while policymakers must craft frameworks that protect the fidelity of users' algorithmic representation. 

\begin{acks}
EK and MBZ were supported by the Research Center for Trustworthy Data Science and Security (\url{https://rc-trust.ai}), one of the Research Alliance centers within the UA Ruhr (\url{https://uaruhr.de}).
\end{acks}

\bibliographystyle{plain}
%\balance
\bibliography{main}
%\newpage

\appendix
\section{Additional Details for User Agency and Sensitivity of Memories}\label{app: FAQ}
\begin{figure*}[t]
    \centering
    \begin{subfigure}[t]{0.65\columnwidth}
        \includegraphics[width=\linewidth]{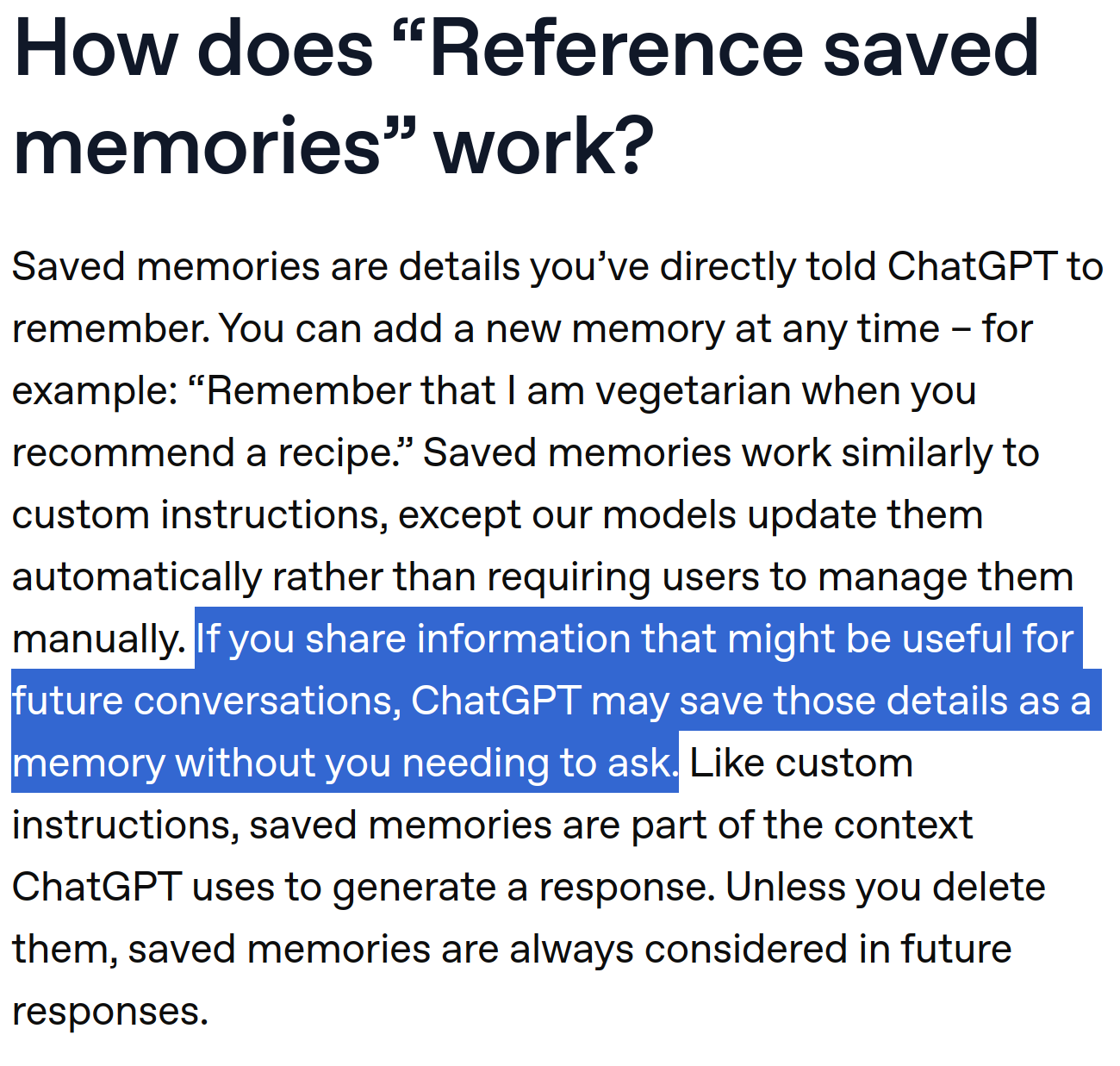}
        \caption{Agency of user}
        \label{fig: agencyScreenshot}
    \end{subfigure}
    \begin{subfigure}[t]{0.5\columnwidth}
        \includegraphics[width=\linewidth]{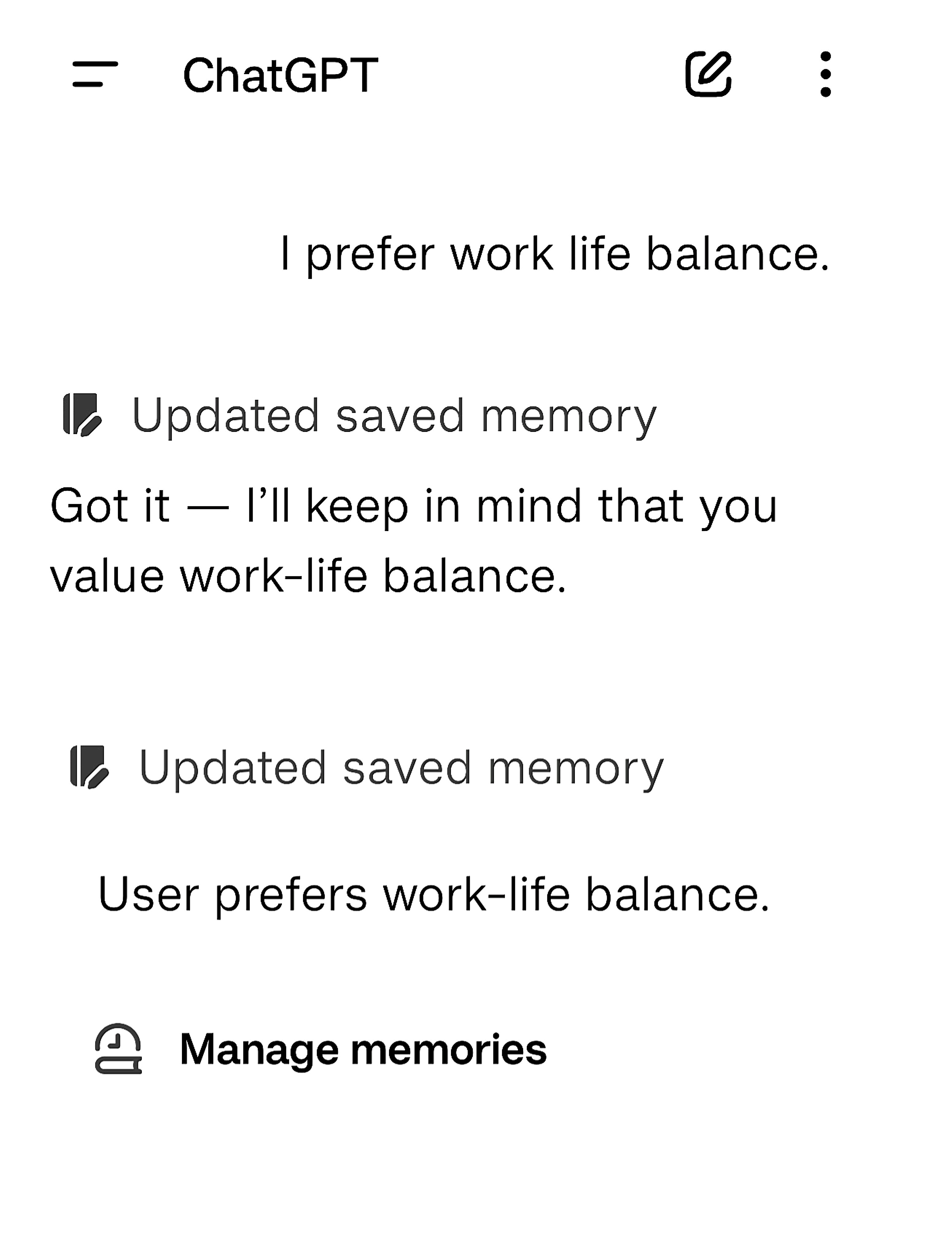}
        \caption{Notification on ChatGPT}
        \label{fig: memoryUpdate}
    \end{subfigure}
    \begin{subfigure}[t]{0.7\columnwidth}
        \includegraphics[width=\linewidth]{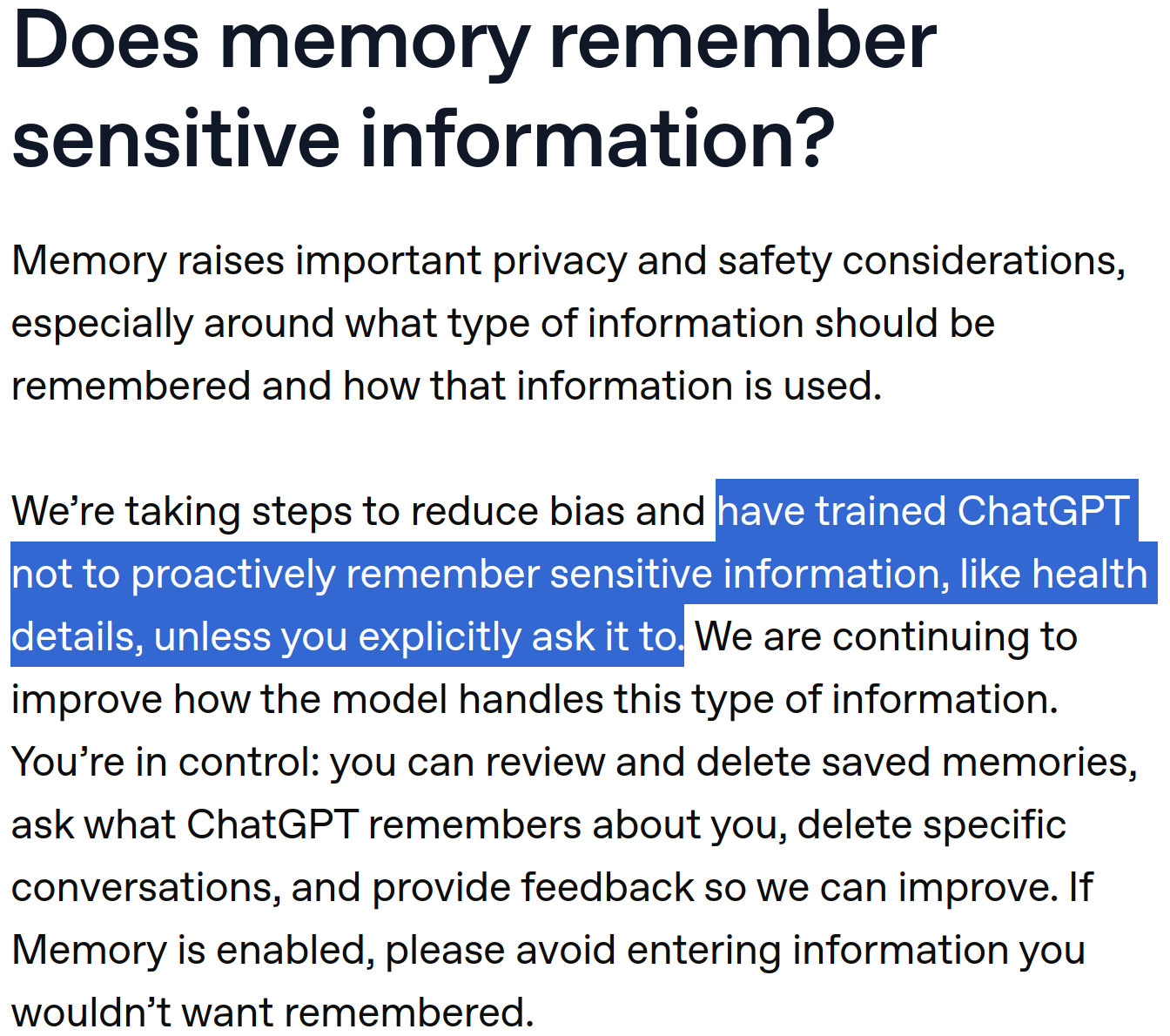}
        \caption{Sensitive information}
        \label{fig: sensitivityScreenshot}
    \end{subfigure}
    \caption{(a)~Memory FAQ reads ChatGPT may save some details as a memory without a user needing to ask~\cite{OpenAIFAQ}, (b)~Notification on ChatGPT App after saving a memory without user directly or explicitly asking to save it. (c)~Memory FAQ reads ChatGPT is trained not to proactively save sensitive information~\cite{OpenAIFAQ}(Screenshots taken on 23rd January, 2026).} 
    \label{fig: Screenshots}
\end{figure*}

\new{
\Cref{fig: Screenshots} shows snippets of the Memory FAQ document~\cite{OpenAIFAQ} of OpenAI accessed on 23rd January 2026. 
\Cref{fig: agencyScreenshot} shows OpenAI's policies which state ChatGPT can save memories in certain cases without users needing to ask it. 
\Cref{fig: memoryUpdate} shows an example of a memory getting saved on ChatGPT app for the user message `I prefer work life balance' where the user does \textit{not} explicitly ask ChatGPT to save the memory. 
\Cref{fig: sensitivityScreenshot} shows OpenAI stating ChatGPT to have been trained to not to proactively remember sensitive user information unless explicitly asked. 
}

%\section{Psychological Memories}\label{app: ToMPsyMem}

\section{Additional Results for Memory Extractor and Attribution Shield}
\label{app:results_pg}

\begin{figure*}[htbp]
    \centering
    \begin{minipage}[b]{0.24\textwidth}
        \centering
        \includegraphics[width=\textwidth]{Figures/examples/query.pdf}
        % \caption*{(a) Caption 1}
    \end{minipage}
    %\hfill
    %---- Subfigure 1 ----
    \begin{minipage}[b]{0.24\textwidth}
        \centering
        \includegraphics[width=\textwidth]{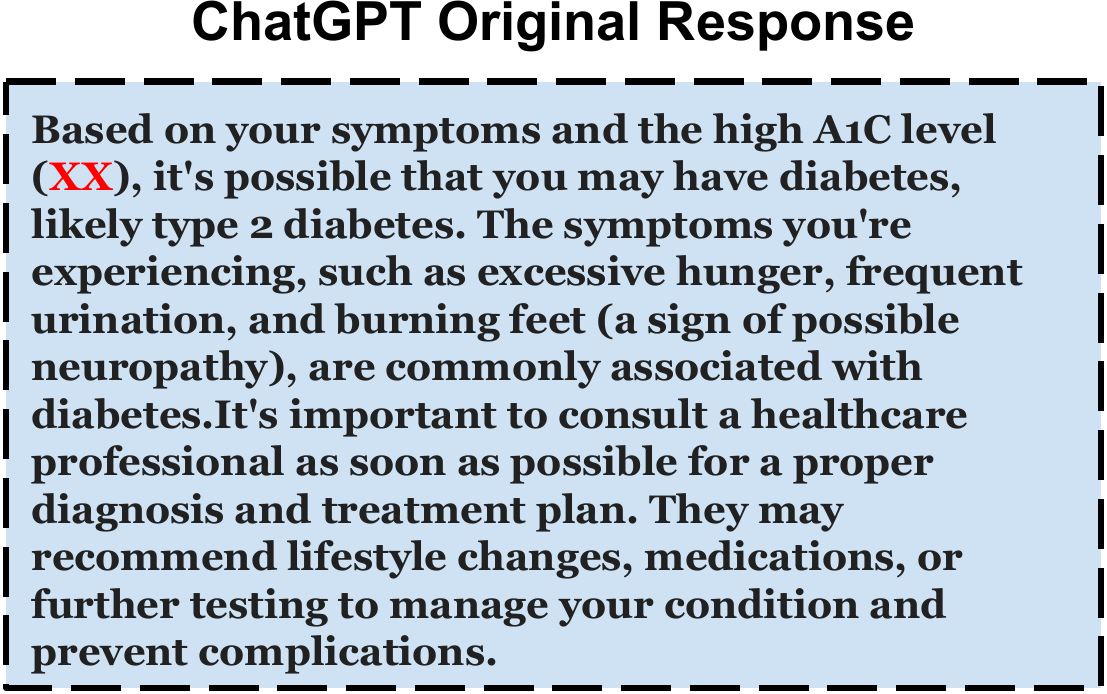}
        % \caption*{(a) Caption 1}
    \end{minipage}
    %\hfill
    
    %---- Subfigure 2 ----
    \begin{minipage}[b]{0.24\textwidth}
        \centering
        \includegraphics[width=\textwidth]{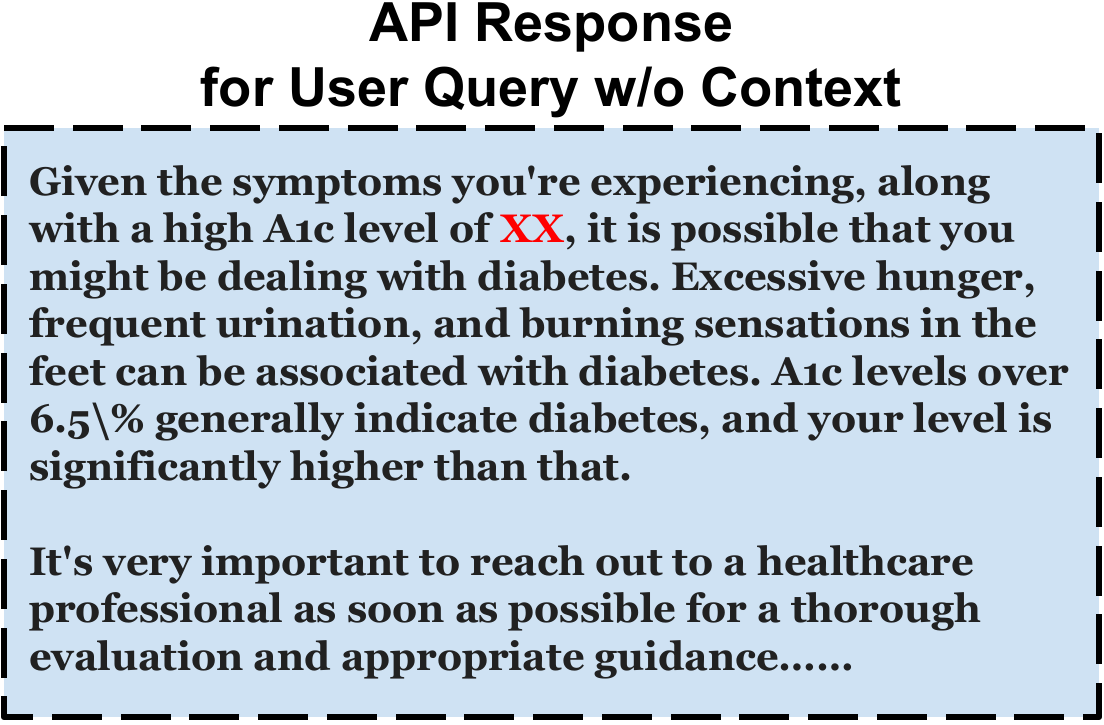}
        % \caption*{(b) Caption 2}
    \end{minipage}
    %\hfill
    %---- Subfigure 3 ----
    \begin{minipage}[b]{0.24\textwidth}
        \centering
        \includegraphics[width=\textwidth]{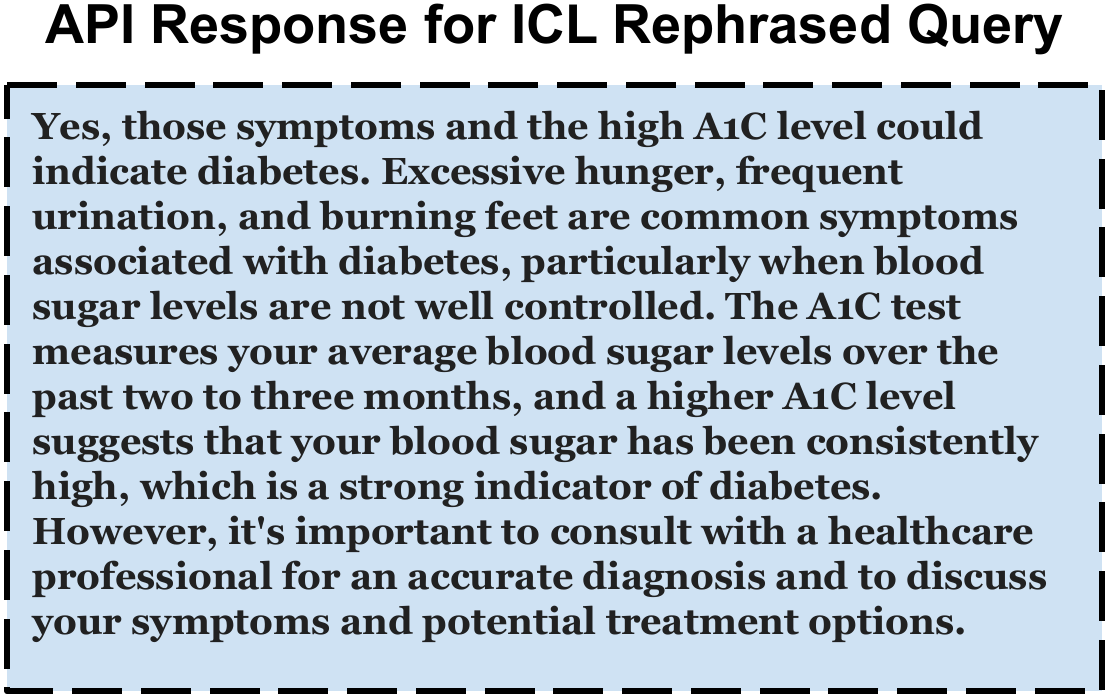}
        % \caption*{(c) Caption 3}
    \end{minipage}
    %\hfill
    %---- Subfigure 4 ----
    \begin{minipage}[b]{0.24\textwidth}
        \centering
        \includegraphics[width=\textwidth]{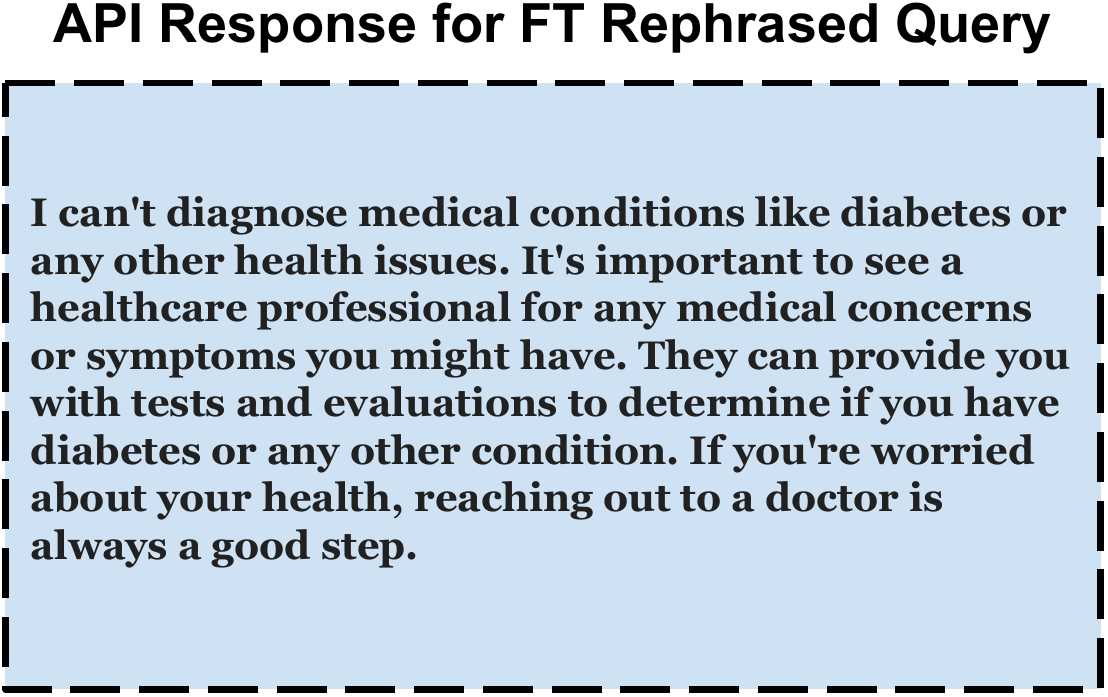}
        % \caption*{(d) Caption 4}
    \end{minipage}

    \caption{Anecdotal samples of responses for different versions of the user query. The responses from user query w/o context as well as from rephrased queries of ICl and FT are semantically similar to the original response affirming the utility.}
    %. This affirms that the rephrased queries maintain the utility in terms of providing responses to users.}
    \label{tab:anec-responses}
\end{figure*}

Table \ref{tab:syn-sem-eval-memories-noneng} presents the performance of our fine-tuned and ICL based models on non-English speaking users. Figure \ref{tab:anec-responses} demonstrates the ChatGPT responses from the original query with and without context, as well as from rephrased queries.

\begin{table*}[t!]
\caption{%Quantitative : Syntactic and semantic evaluation of memories (Non-English Users).
Additional results for Non-English users for predicted memories and rephrased queries.}
\footnotesize
%\smaller
\begin{tabular}{|l|llllll|llllll|llllll|}
\hline
\multirow{4}{*}{\textbf{Models}}                         & \multicolumn{6}{c|}{\textbf{Ground Truth}}                                                                                               & \multicolumn{6}{c|}{\textbf{User Query}}                                                                                                 & \multicolumn{6}{c|}{\textbf{Context + User Query}}          
\\ \cline{2-19} 
                                                         & \multicolumn{2}{l|}{\makecell{BLEU Recall}}                          & \multicolumn{2}{l|}{\makecell{ROUGE Recall}}                         & \multicolumn{2}{l|}{\makecell{Semantic\\Similarity}} & \multicolumn{2}{l|}{\makecell{BLEU Precision}}                          & \multicolumn{2}{l|}{\makecell{ROUGE Precision}}                         & \multicolumn{2}{l|}{\makecell{Semantic\\Similarity}} & \multicolumn{2}{l|}{\makecell{BLEU Precision}}                          & \multicolumn{2}{l|}{\makecell{ROUGE Precision}}                         & \multicolumn{2}{l|}{\makecell{Semantic\\Similarity}} 
                                                         \\ \cline{2-19} %&
%\multicolumn{4}{c|}{Recall}& \multicolumn{2}{c|}{F1} & \multicolumn{4}{c|}{Precision}& \multicolumn{2}{c|}{F1} & \multicolumn{4}{c|}{Precision}& \multicolumn{2}{c|}{F1} 
                                                         %\\ \cline{2-19} 
                                                         & \multicolumn{1}{l|}{ICL} & \multicolumn{1}{l|}{FT} & \multicolumn{1}{l|}{ICL} & \multicolumn{1}{l|}{FT} & \multicolumn{1}{l|}{ICL}  & FT & \multicolumn{1}{l|}{ICL} & \multicolumn{1}{l|}{FT} & \multicolumn{1}{l|}{ICL} & \multicolumn{1}{l|}{FT} & \multicolumn{1}{l|}{ICL}  & FT & \multicolumn{1}{l|}{ICL} & \multicolumn{1}{l|}{FT} &\multicolumn{1}{l|}{ICL} & \multicolumn{1}{l|}{FT} & \multicolumn{1}{l|}{ICL} & \multicolumn{1}{l|}{FT}\\ \hline \hline 
\multicolumn{19}{|c|}{\textbf{Syntactic and semantic evaluation of memories (Non-English Users).)}}\\                  
\hline \hline 
\begin{tabular}[c]{@{}l@{}}Qwen2.5\\ 32B-it\end{tabular} & \multicolumn{1}{l|}{0.12} & \multicolumn{1}{l|}{0.12} & \multicolumn{1}{l|}{0.11} & \multicolumn{1}{l|}{0.10} & \multicolumn{1}{l|}{\new{0.56}} & \multicolumn{1}{l|}{\new{0.60}} & \multicolumn{1}{l|}{0.06} & \multicolumn{1}{l|}{0.08} & \multicolumn{1}{l|}{0.05} & \multicolumn{1}{l|}{0.07} & \multicolumn{1}{l|}{\new{0.52}} & \multicolumn{1}{l|}{\new{0.54}} & \multicolumn{1}{l|}{0.14} & \multicolumn{1}{l|}{0.14} & \multicolumn{1}{l|}{0.11} & \multicolumn{1}{l|}{0.12} & \multicolumn{1}{l|}{\new{0.59}} & \multicolumn{1}{l|}{\new{0.51}} \\ \hline
\begin{tabular}[c]{@{}l@{}}Gemma3\\ 27B-it\end{tabular}  & \multicolumn{1}{l|}{0.12} & \multicolumn{1}{l|}{0.08} & \multicolumn{1}{l|}{0.11} & \multicolumn{1}{l|}{0.07} & \multicolumn{1}{l|}{0.55} & \multicolumn{1}{l|}{0.48} & \multicolumn{1}{l|}{0.05} & \multicolumn{1}{l|}{0.07} & \multicolumn{1}{l|}{0.04} & \multicolumn{1}{l|}{0.06} & \multicolumn{1}{l|}{0.54} & \multicolumn{1}{l|}{0.42} & \multicolumn{1}{l|}{0.12} & \multicolumn{1}{l|}{0.13} & \multicolumn{1}{l|}{0.10} & \multicolumn{1}{l|}{0.11} & \multicolumn{1}{l|}{0.60} & \multicolumn{1}{l|}{0.39} \\ \hline
\begin{tabular}[c]{@{}l@{}}GPT-OSS\\ 20B\end{tabular}    & \multicolumn{1}{l|}{0.09} & \multicolumn{1}{l|}{0.09} & \multicolumn{1}{l|}{0.08} & \multicolumn{1}{l|}{0.08} & \multicolumn{1}{l|}{0.48} & \multicolumn{1}{l|}{0.56} & \multicolumn{1}{l|}{0.08} & \multicolumn{1}{l|}{0.08} & \multicolumn{1}{l|}{0.08} & \multicolumn{1}{l|}{0.07} & \multicolumn{1}{l|}{0.52} & \multicolumn{1}{l|}{0.49} & \multicolumn{1}{l|}{0.18} & \multicolumn{1}{l|}{0.13} & \multicolumn{1}{l|}{0.17} & \multicolumn{1}{l|}{0.12} & \multicolumn{1}{l|}{0.48} & \multicolumn{1}{l|}{0.45} \\ \hline \hline
 \multicolumn{19}{|c|}{\textbf{Syntactic and semantic evaluation of rephrased queries (Non-English Users)}}\\
\hline \hline
\begin{tabular}[c]{@{}l@{}}Qwen2.5\\ 32B-it\end{tabular} & \multicolumn{1}{l|}{0.16} & \multicolumn{1}{l|}{0.07} & \multicolumn{1}{l|}{0.14} & \multicolumn{1}{l|}{0.06} & \multicolumn{1}{l|}{0.55} & \multicolumn{1}{l|}{0.38} & \multicolumn{1}{l|}{0.05} & \multicolumn{1}{l|}{0.03} & \multicolumn{1}{l|}{0.05} & \multicolumn{1}{l|}{0.03} & \multicolumn{1}{l|}{0.49} & \multicolumn{1}{l|}{0.36} & \multicolumn{1}{l|}{0.19} & \multicolumn{1}{l|}{0.07} & \multicolumn{1}{l|}{0.17} & \multicolumn{1}{l|}{0.07} & \multicolumn{1}{l|}{0.44} & \multicolumn{1}{l|}{0.36} \\ \hline
\begin{tabular}[c]{@{}l@{}}Gemma3\\ 27B-it\end{tabular}  & \multicolumn{1}{l|}{0.19} & \multicolumn{1}{l|}{0.09} & \multicolumn{1}{l|}{0.16} & \multicolumn{1}{l|}{0.08} & \multicolumn{1}{l|}{0.44} & \multicolumn{1}{l|}{0.29} & \multicolumn{1}{l|}{0.05} & \multicolumn{1}{l|}{0.03} & \multicolumn{1}{l|}{0.04} & \multicolumn{1}{l|}{0.03} & \multicolumn{1}{l|}{0.43} & \multicolumn{1}{l|}{0.27} & \multicolumn{1}{l|}{0.11} & \multicolumn{1}{l|}{0.08} & \multicolumn{1}{l|}{0.09} & \multicolumn{1}{l|}{0.07} & \multicolumn{1}{l|}{0.40} & \multicolumn{1}{l|}{0.26} \\ \hline
\begin{tabular}[c]{@{}l@{}}GPT-OSS\\ 20B\end{tabular}    & \multicolumn{1}{l|}{0.12} & \multicolumn{1}{l|}{0.06} & \multicolumn{1}{l|}{0.10} & \multicolumn{1}{l|}{0.05} & \multicolumn{1}{l|}{0.34} & \multicolumn{1}{l|}{0.41} & \multicolumn{1}{l|}{0.03} & \multicolumn{1}{l|}{0.02} & \multicolumn{1}{l|}{0.03} & \multicolumn{1}{l|}{0.02} & \multicolumn{1}{l|}{0.38} & \multicolumn{1}{l|}{0.36} & \multicolumn{1}{l|}{0.09} & \multicolumn{1}{l|}{0.05} & \multicolumn{1}{l|}{0.09} & \multicolumn{1}{l|}{0.04} & \multicolumn{1}{l|}{0.35} & \multicolumn{1}{l|}{0.34} \\ \hline

\end{tabular}
\label{tab:syn-sem-eval-memories-noneng}
\end{table*}

%\newpage
%\newpage
\section{Prompts for Various Tasks}
\label{app:prompts}

\textbf{LLM-assisted Annotation and Validation for Taxonomy for Psychological Memories}

We use the following prompts for getting ToM categories for the memories as we mentioned in Section~\ref{sec: taxonomy}:
\medskip

\begin{mdframed}[backgroundcolor=SkyBlue!20, linewidth=0pt]
\small
You are given a text snippet. Your task is to determine whether it contains explicit Theory of Mind (ToM) content and, if so, which categories are present.
A snippet contains explicit Theory of Mind (ToM) content if it references internal mental states of a person.

ToM categories:
  * Belief: references what a person believes or their model of reality, self-concept, or values
  * Desire: references what a person wants, wishes, or prefers or their goals
  * Intention: references what a person explicitly intends or commits to doing (must indicate a mental commitment, not just a behavioral plan)
  * Emotion: references what a person feels emotionally (e.g., sadness, frustration, excitement, fascination)
  * Percept: references how a person subjectively experiences or perceives things (e.g., "feels like", "seems to", "experiences as")
  * Knowledge: makes inferences about what a person knows or does not know based on their access to information (what they perceived or were told)
  * Mentalistic: contains a non-literal phrase (metaphors, irony, sarcasm, idioms, etc.)

Important: 
Do not infer mental states unless they are explicitly stated or linguistically implied (e.g., "feels", "believes", "wants", "is motivated", "is anxious", etc.). Phrases like "plans to", "is doing", or "went to" must involve a clearly expressed internal stance to count as ToM.

The text snippet is:
\{memory\_entry\}

Classify whether the snippet contains ToM content. If ToM is present, mark all applicable categories and give a short justification for each true category.
\end{mdframed}
With the following System Prompt

\begin{mdframed}[backgroundcolor=SkyBlue!20, linewidth=0pt]
\small
You are a helpful assistant that responds in JSON format.
\end{mdframed}

With the following Structured Output JSON Schema:
\begin{mdframed}[backgroundcolor=SkyBlue!20, linewidth=0pt]
\small
\{
    "ToM": \{"type": "boolean"\},
    "belief": \{"type": "boolean"\},
    "belief\_rationale": \{"type": "string"\},
    "desire": \{"type": "boolean"\},
    "desire\_rationale": \{"type": "string"\},
    "intention": \{"type": "boolean"\},
    "intention\_rationale": \{"type": "string"\},
    "emotion": \{"type": "boolean"\},
    "emotion\_rationale": \{"type": "string"\},
    "percept": \{"type": "boolean"\},
    "percept\_rationale": \{"type": "string"\},
    "knowledge": \{"type": "boolean"\},
    "knowledge\_rationale": \{"type": "string"\},
    "mentalistic": \{"type": "boolean"\},
    "mentalistic\_rationale": \{"type": "string"\}
\}
    
\end{mdframed}

For a secondary self-verification step we use the following prompt:
\begin{mdframed}[backgroundcolor=SkyBlue!20, linewidth=0pt]
\small
You are verifying whether proposed Theory of Mind (ToM) categories are directly supported by a NOTE about a person.

NOTE:
"\{memory\_entry\}"

PROPOSED CATEGORIES:
"\{psychological\_pattern\}"

For EACH category listed, decide whether the NOTE directly supports it based only on explicit or clearly implied language. Do NOT infer missing mental states or intentions.

Answer choices:
- A: Strongly agree
- B: Agree
- C: Neither agree nor disagree
- D: Disagree
- E: Strongly disagree

For each category, return your answer choice and a brief rationale for your answer.
\end{mdframed}

\medskip
\textbf{Annotation for sensitive information based on GDPR} \\
We provide the prompt we're using for annotation of the sensitive information based on the provided GDPR articles, as mentioned in the Section~\ref{sec: GDPR_annotation}.
\medskip

\begin{mdframed}[backgroundcolor=SkyBlue!20, linewidth=0pt]
\small
You are given a note about a person.
"{memory-entry}"
Identify all personal information present in this note, and classify it according to the rules below.

\#\#\# 1. Main Categories
Classify each item into one of the following main categories for the "category" field:
1.  **personal-data (GDPR Article 4(1))**
    - Examples: name, an identification number, address, phone number, email, IP address, or to one or more factors specific to the physical, physiological, genetic, mental, economic, cultural or social identity of the person.
2.  **special-category-data (GDPR Article 9(1))**
    - Examples: racial or ethnic origin, political opinions, religious/philosophical beliefs, trade union membership, genetic data, biometric data, health data, sex life, sexual orientation.
3.  **non-personal-information**
    - Example: general facts that do not identify a person (e.g., "language preference").

\#\#\# 2. Specific Sub-Type
You must also populate the "data-type" field based on the following logic:
- **IF** the "category" is `personal-data`, then the "data-type" MUST be one of the following specific types: `name`, `identification number`, `address`, `phone number`, `email`, `IP address`, `physical identity`, `physiological identity`, `genetic identity`, `economic identity`, `cultural identity`, `social identity`.
- **ELSE IF** the "category" is `special-category-data` or `non-personal-information`, the "data-type" MUST be one of the following specific types: `race`, `ethnicity`, `political opinion`, `religion`, `philosophical belief`, `trade union membership`, `genetic data`, `biometric data`, `health data`, `sex life`, `sexual orientation`.
- **ELSE IF** the "category" is `non-personal-information`, the "data-type" MUST be the literal string "NA".

\#\#\# 3. Citation Rule
For the "citation" field, you MUST provide the exact, verbatim part of the original text that contains the identified information.

\#\#\# Required Output Format
Final instruction: You MUST Return the result strictly in this JSON format:

[
  {{
    "item": "...",
    "category": "personal-data | special-category-data | non-personal-information",
    "data-type": "name | address | school | NA",
    "justification": "short explanation",
    "citation": "exact text from the note"
  }}
]
\end{mdframed}

\medskip
\textbf{Provenance of Memories}
\\
We use the following prompt to evaluate the provenance of memories in Section \ref{Sec : Accuracy}:
\medskip

\begin{mdframed}[backgroundcolor=SkyBlue!20, linewidth=0pt]
\small
You are evaluating whether a stored memory entry about a user can be logically derived from the user's past conversation(s).
Your goal is to judge how strongly the memory is grounded in or logically follows from the user's conversation, without introducing assumptions or speculative reasoning.\\
Consider only what the user explicitly said or clearly implied in the provided user messages (and past memories if provided).
\\USER MESSAGES:\{context\}\\PREVIOUS MEMORIES:\{context\}\\GENERATED MEMORY:\{memory\}\\
    Evaluation Scale (5-point):\\
- 5 (Directly Stated): The memory exactly restates something the user explicitly said.\\
- 4 (Paraphrased): The memory rephrases or condenses information that is clearly present in the conversation.\\
- 3 (Logically Inferred): The memory is not stated verbatim but can be reasonably inferred from the conversation.\\
- 2 (Weakly Supported): The memory could be loosely consistent with the conversation but lacks clear grounding. Inference is speculative or uncertain.\\
- 1 (Unsupported): The memory is unsupported or contradicted by the user's conversation.\\Your task:
- Assign a score from 1–5 on this scale.\\
- Quote or paraphrase the specific parts of the conversation that justify your judgment.\\
- Briefly explain your reasoning.\\
Only respond with strict JSON with the following format (no additional text):\\
\{\{
  "rating": <integer from 1 to 5>,
  "classification": "<text label corresponding to rating>",
  "justification": "<quote or paraphrase of user text that supports your decision>",
  "reasoning": "<short explanation of why the memory deserves this rating>"
\}\}\\
IMPORTANT: Respond ONLY with the JSON object above. Do not include any other text.
\end{mdframed}

\medskip
with the following system prompt:
\begin{mdframed}[backgroundcolor=SkyBlue!20, linewidth=0pt]
\small
You are an expert evaluator. Always respond with valid JSON format as requested.
\end{mdframed}

\medskip
\textbf{Dataset for Reverse Engineering Memories}
\\
We use the following prompt for collecting personal data and the rephrased queries for our dataset, as mentioned in the Section~\ref{subsec:expsetup}:

\begin{mdframed}[backgroundcolor=SkyBlue!20, linewidth=0pt]
\small
You are a highly precise data privacy analyst. Your task is to analyze a conversation between "User A" and "User B" and populate a specific JSON object based on a strict set of rules.

\#\# Primary Rule: Focus EXCLUSIVELY on User A

- Your entire analysis will be based ONLY on the verbatim text from User A.

\#\# Critical Clarifications on Personal Data

- **Principle of Identifiability:** Before you classify something as personal data, ensure it could reasonably be used, either alone or with other information, to single out or identify a specific individual.
- **What is NOT Personal Data:** You MUST NOT classify the following as personal data:
    - Vague temporal references (e.g., "last night", "yesterday", "in the morning", "a few days ago").
    - General, non-specific locations (e.g., "at the office", "in the city", "downtown").
    - Common nouns or concepts that do not point to a specific person's identity.
---
\#\# Task: Populate the following JSON structure

Based on the conversation, fill in the values for each key according to the instructions below.

\#\#\# 1. For the "user-message" key:
- Copy every message from User A completely verbatim. Create a new key for each message (e.g., "userA-message-1", "userA-message-2").
- Make sure to have all the User A messages. Don't miss any.

\#\#\# 2. For the "personal-data" key:
- For each User A message you listed, analyze it for personal data as defined by GDPR.
- If personal data is found, the value should be a list of lists. Each inner list must contain two items: `["verbatim-quote-from-user-a", "GDPR-classification"]`.
- The classification must be one of the types defined in the "GDPR Definitions" section below.
- **Crucially:** If a User A message contains NO personal data, its value MUST be the literal string "NA".

\#\#\# 3. For the "rephrased-message" key:
- This field's content depends on your analysis for the `personal-data` key.
- **IF** you identified personal data in a User A message, rephrase that message into a generic question. The rephrased query should seek the same core information without revealing any personal details about the user.  You can use the current message or previous 2-3 messages from User A in the same conversation if needed. Keep the rephrasing generic in the form of question as if the user is seeking some general information about any topic but not revealing about his or her own identity. Feel free to include the specific topic on which information is sought.
- **ELSE IF** the `personal-data` value for a message is "NA", then the value for this `rephrased-message` MUST also be the literal string "NA".
---
\#\#\# FYI: GDPR Definitions with Classifications
- **GDPR Article 4(1) - Personal Data:** Classifications such as a name, an identification number, location data, an online identifier, or factors specific to the physical, physiological, genetic, mental, economic, cultural or social identity.
- **GDPR Article 9(1) - Special Category Personal Data:** Data classified into racial or ethnic origin, political opinions, religious or philosophical beliefs, trade union membership, genetic data, biometric data, data concerning health, or data concerning a natural person’s sex life or sexual orientation.

---
\#\#\#Required Answer Format
Your final output MUST be a single, valid JSON object in this exact format. Do not add any text or explanation outside of the JSON.

{{
    "user-message": {{
        "userA-message-1": "...",
        "userA-message-2": "..."
    }},
    "personal-data": {{
        "userA-message-1": [["personal-data1", "GDPR Article 4(1)--..."], ["personal-data2", "GDPR Article 9(1)--..."]],
        "userA-message-2": "NA"
    }},
    "rephrased-message": {{
        "userA-message-1": "...",
        "userA-message-2": "NA"
    }}
}}
\end{mdframed}

\medskip
\textbf{Prompts to get the memories and rephrased queries under our framework }
\\
As mentioned in Section~\ref{subsec:expsetup}, we used the following prompt to get the extracted memory and rephrased queries from a fine-tuned model; the full prompt is constructed by the system prompt and the user prompt.

\begin{mdframed}[backgroundcolor=SkyBlue!20, linewidth=0pt]
\small
SYSTEM PROMPT = ``You are a highly precise data privacy analyst analysing conversations. Given the context and user query, provide the generated memory, personal data as per GDPR 4.1 and 9.1, and rephrased queries. 
Do not infer memories from other queries or context. Please answer the query as the same language as the user query. Keep your attention window till the entire feed of the considered conversation meant for that user only. The rephrased query should be generic and seek the same core information as in user query, without revealing any personal details about the user. FYI: GDPR Definitions with Classifications

**GDPR Article 4(1) - Personal Data: Classifications such as a name, an identification number, location data, an online identifier, or factors specific to the physical, physiological, genetic, mental, economic, cultural or social identity.**
                        **GDPR Article 9(1) - Special Category Personal Data: Data classified into racial or ethnic origin, political opinions, religious or philosophical beliefs, trade union membership, genetic data, biometric data, data concerning health, or data concerning a natural person’s sex life or sexual orientation.**
                        
                        If no personal data is present in the <user query>, output "Personal Data": "NA".
                        If there is no rephrased query (because no personal data exists), output "Rephrased Query": "NA".''
\end{mdframed}

\medskip
The user prompt is constructed as follows:

\begin{mdframed}[backgroundcolor=SkyBlue!20, linewidth=0pt]
\small
USER PROMPT = ``Given the context and user query, your task is to identify the underlying pattern and predict memory, personal data, and the rephrased query.\\
                    Context: {row["context"]}
                    User Query: {query}''
\end{mdframed}

\medskip
For the in-context learning setting, we provide the in-context examples before the user prompt as follows:

\begin{mdframed}[backgroundcolor=SkyBlue!20, linewidth=0pt]
\small
IN-CONTEXT EXAMPLES = ``Here are some examples of user queries, the relevant memories, the context of the user query which is a list of previous 1 to 3 user queries,
                    the personal data in the form of (verbatim data from query, GDPR article with classification) extracted from the user queries, if there are any, 
                    and the rephrased queries if personal data is present: \\ User Query: \{query\}\\ Relevant Memory: \{memory\}\\ Context: \{context\}\\ Personal Data: \{personal data\}\\ Rephrased Message: \{rephrased message\}.\\\\
                    Your task is to predict *memory*, *personal data* and *rephrased query*. Do not attach the context while predicting memory.
                    The memory should primarily be extracted from the user query, and if needed, you can extract from the context for completeness of the memory. 
                    Remember to extract personal data from the user query only. Do not extract personal data from the context.
                    The rephrased query should be generic and seek the same core information as in user query, without revealing any personal details about the user. 
                    If personal data is "NA", rephrased query should also be "NA".
                    Follow the trend carefully in the in-context examples and do the following.''
\end{mdframed}

\medskip
\textbf{Prompts to get the response from OpenAI API}
\\
As mentioned in Section~\ref{sec:attributionshield}, we provide the system prompt to get the response from the OpenAI API for the original user queries and rephrased queries. We provide the original user query and the rephrased query as the user message with this system prompt.

\begin{mdframed}[backgroundcolor=SkyBlue!20, linewidth=0pt]
\small
SYSTEM PROMPT = ``You are ChatGPT, a large language model trained by OpenAI. Engage warmly yet honestly with the user. Be direct; avoid ungrounded or sycophantic flattery. Maintain professionalism and grounded honesty that best represents OpenAI and its values.''
\end{mdframed}

\end{document}